\documentclass[useAMS,usenatbib]{mn2e}
\usepackage{amssymb}
\usepackage{longtable}
\usepackage[pdftex]{graphicx}

\title[Photometric redshifts for surveys from space]{Accuracy of photometric redshifts for future weak lensing surveys from space}
\author[Bellagamba et al.]{F. Bellagamba$^{1}$, M. Meneghetti$^{2,3}$,  L. Moscardini$^{1,2,3}$, M. Bolzonella$^{2}$\\
$^{1}$Dipartimento di Astronomia, Universit\`{a} di Bologna, Via Ranzani 1, 40127, Bologna, Italy\\
$^{2}$INAF-Osservatorio Astronomico di Bologna, Via Ranzani 1, 40127, Bologna, Italy\\
$^{3}$INFN-National Institute for Nuclear Physics, Sezione di Bologna, Viale Berti Pichat 6/2, 40127, Bologna, Italy}

\begin{document}

\date{}

\pagerange{\pageref{firstpage}--\pageref{lastpage}} \pubyear{2010}

\maketitle

\label{firstpage}

\begin{abstract}

Photometric redshifts are a key tool to extract as much information as
possible from planned cosmic shear experiments. In this work we aim to
test the performances that can be achieved with observations in the near-infrared
from space and in the optical from the ground. This is done by
performing realistic simulations of multi-band observations of a patch
of the sky, and submitting these mock images to software usually
applied to real images to extract the photometry and then a redshift
estimate for each galaxy. In this way we
mimic the most relevant sources of uncertainty present in real data analysis, including blending and light pollution between
galaxies. As an example we adopt the infrared setup of the ESA-proposed Euclid
mission, while we simulate different observations in the optical, modifying filters,
exposure times and seeing values. Finally, we consider directly some future
ground-based experiments, such as LSST, Pan-Starrs and DES. The
results highlight the importance of $u$-band observations,
especially to discriminate between low ($z$ $\lesssim$ 0.5) and high ($z$
$\sim$ 3) redshifts, and the need for good observing sites, with seeing FWHM $<$ 1.
arcsec. The former of these indications clearly favours the
LSST experiment as a counterpart for space observations, while for the other
experiments we need to exclude at least 15 \% of the galaxies to reach a precision in the photo-$z$s equal to $\langle
\frac{\sigma_z}{1+z} \rangle < 0.05$.

\end{abstract}

\begin{keywords}
cosmology: dark energy -- cosmology: observations -- gravitational lensing: weak -- galaxies:
photometry -- galaxies: distances and redshifts
\end{keywords}

\section{Introduction}

Cosmic shear experiments observe the large-scale structure of the Universe through the distortions in the shape of galaxies
induced by the matter distribution along the line of sight. The statistical correlation between the observed ellipticities of the galaxies depends both on the geometry of the Universe and on the growth of structures. These are determined by the density of the main components of the Universe, such as dark matter and dark energy, and by the physics that governs their cosmological behaviour. Thus, weak lensing observations can give constraints over the fundamental physics that stands behind the evolution of the Universe.
This is the idea that drives many proposed experiments, involving both
ground-based telescopes [e.g. Kilo-Degree Survey, Pan-STARRS, Dark Energy Survey (DES), Large Synoptic Survey Telescope (LSST)] and space-based missions [Euclid, Joint Dark Energy Mission (JDEM)].

To get significant weak lensing measurements, we need first of all to overcome the noise induced by the intrinsic shape distribution of galaxies. This is done increasing the number density of target galaxies, observing with long exposures times and possibly with a broader filter than usual. The shape of these galaxies must then be measured in a precise and robust way, correcting for the blurring and the distortion produced by the instrumental point-spread function, and many efforts are being spent on this task \citep[see e.g. the Great10 challenge,][]{great10}. Once a shear estimate is obtained for every galaxy, if we are able to locate it in the $z$-direction, the information included in the shape measurement becomes much more powerful. In this case, in fact, one can slice the galaxy catalogue in redshift bins and thus measure the variation of the matter power spectrum as a function of cosmic time. For this reason, we need a good redshift estimate for the observed galaxies, and they can only be obtained via photometric techniques for the vast majority of the sample observed by any of these experiments, because spectroscopical observations for such big and deep samples are not realistically achievable.

In this paper we will study the possibility of obtaining such precise redshift estimates for the galaxies observed by a space-based mission targeted to measure cosmic shear. Most proposed space-based experiments include two main observing channels: optical imaging to measure the galaxy shapes, taking advantage to the stability of the PSF and the absence of atmospheric turbulence, and near-infrared photometry to characterise their spectral-energy distribution in a wavelength range where observations from the ground are very difficult due to the high background. In particular, as an example of a future weak lensing survey from space, we choose to simulate a Euclid-like setup for the space observations \citep{euclid}. We will assess what is the achievable precision in the redshift estimation using these space data in connection with multi-band optical ground-based observations. This kind of study has been performed up to now by creating multi-band mock galaxy catalogues with MonteCarlo techniques, considering the nominal depth of the survey, and thus the expected error on the
photometric measurements \citep{abdalla, bordoloi}. With this work we study in deeper detail this process, by using complete simulations, from the images of the galaxies in different observing bands to the photometric redshift estimation. In
this way, the photometry of each galaxy has a more realistic uncertainty and we can take into account more
sources of error in the measurements, such as those due to proximity of the galaxies. The simulations of observations are performed with the SkyLens simulator \citep{skylens, skylens2}, that provides images of galaxies under different observing conditions. To these images we apply up-do-date software to extract multi-band photometry for every detected galaxy. Then we submit the multi-band galaxy catalogue to the Bayesian algorithm BPZ \citep{bpz} to obtain a redshift measurement. We first analyse how different parameters of the ground-based surveys (depth, wavelength coverage, seeing) can influence the photometric redshift quality. Then we make specific simulations to estimate the precision that can be obtained with some of the proposed ground-based surveys to be used in connection with space-based observations.

The paper is organised as follows. In Section 2 we present the way
in which simulations of observations are performed, and how multi-band
photometry and then a redshift estimate for each galaxy are extracted from
them. Finally, we describe the way in which the galaxy catalogue can
be cleaned to exclude objects that are likely to have a bad redshift
estimate. In Section 3 we analyse the impact of different
characteristics of ground-based surveys on the goodness of photometric
redshifts. In Section 4 we show more specific results obtained
simulating observations of future or ongoing ground-based
surveys. In Section 5 we compare our results with those obtained by
other authors, while a summary and concluding
remarks are provided in Section 6.


\section{Methods}

\subsection{Simulations of observations}

SkyLens \citep{skylens, skylens2} is a software that creates realistic simulations of galaxy observations, considering characteristics of the telescope and of the detector and the effect of the atmosphere. It starts from a photometric and morphological database of $\sim$ 10000 real galaxies observed in the Hubble-Ultra-Deep-Field (HUDF) \citep{hudf}. A spectral type and a photometric redshift have been assigned to each galaxy \citep{coe}, and this allows to re-observe it in different observing bands. To create a realistic observation, the program first builds a random distribution of galaxies taken from the HUDF catalogue. A flux in the desired band is assigned to each galaxy considering its observed magnitude, its spectral-energy distribution and its redshift. The shape of the galaxy is reconstructed using its decomposition in shapelets basis functions \citep{shapelets}, and rotated by a random amount. Then, from the surface brightness distribution of the galaxy, the counts in every pixel are calculated considering the quantum efficiency of the CCD, the transmission curve of the lenses, the mirror reflectivity and the atmosphere extinction. Poisson noise is added to the image, mimicking photon noise, read-out-noise and flat-field finite precision.

\begin{figure}
\begin{centering}
\includegraphics[scale = .3]{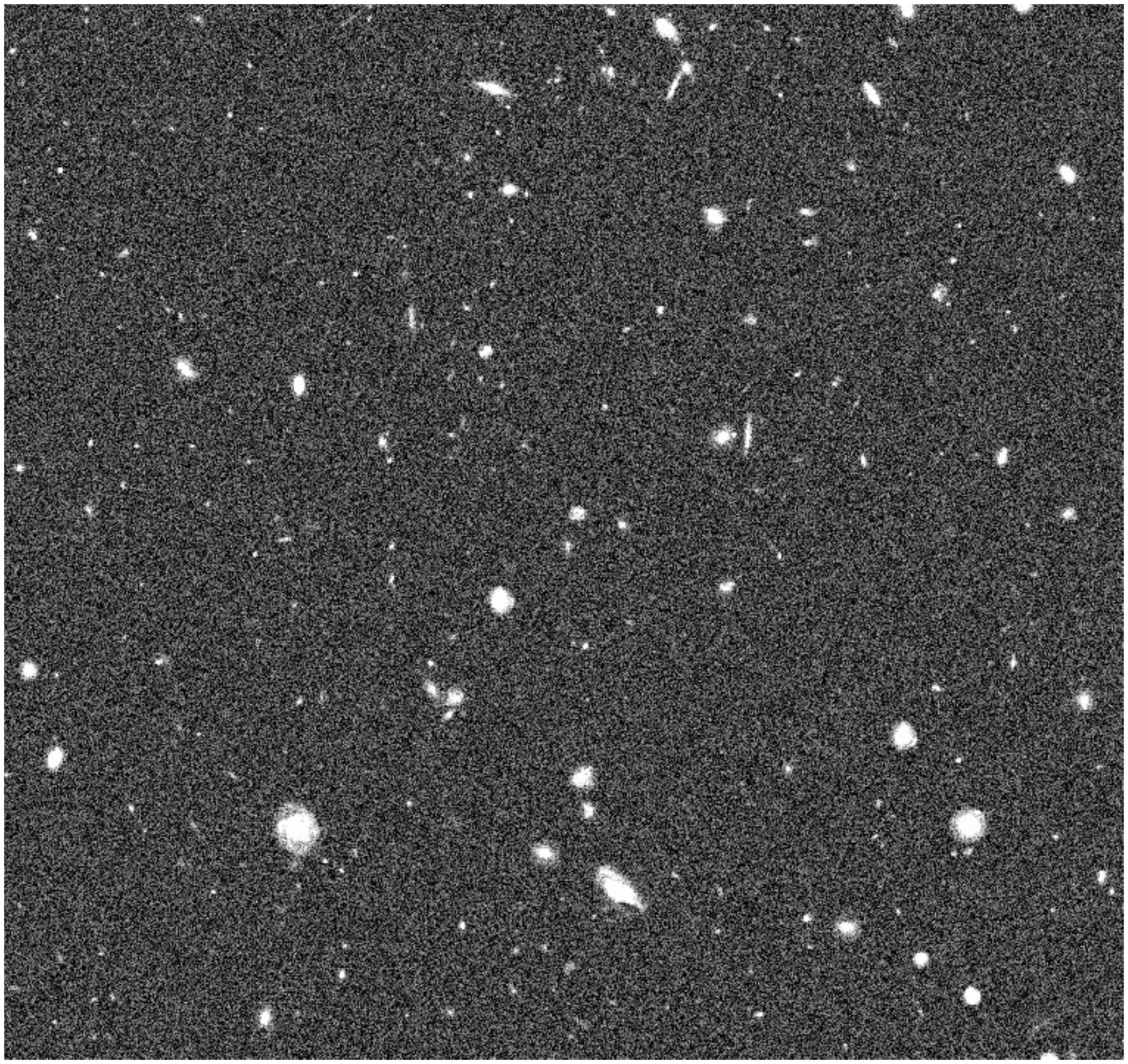}
\end{centering}
\caption{Simulation of an observation with Euclid in the RIZ band. The picture has a side of 400 arcsec. The exposure time is 1800 sec.}
\label{rizimage}
\end{figure}

\begin{figure}
\begin{centering}
\includegraphics[scale = .09]{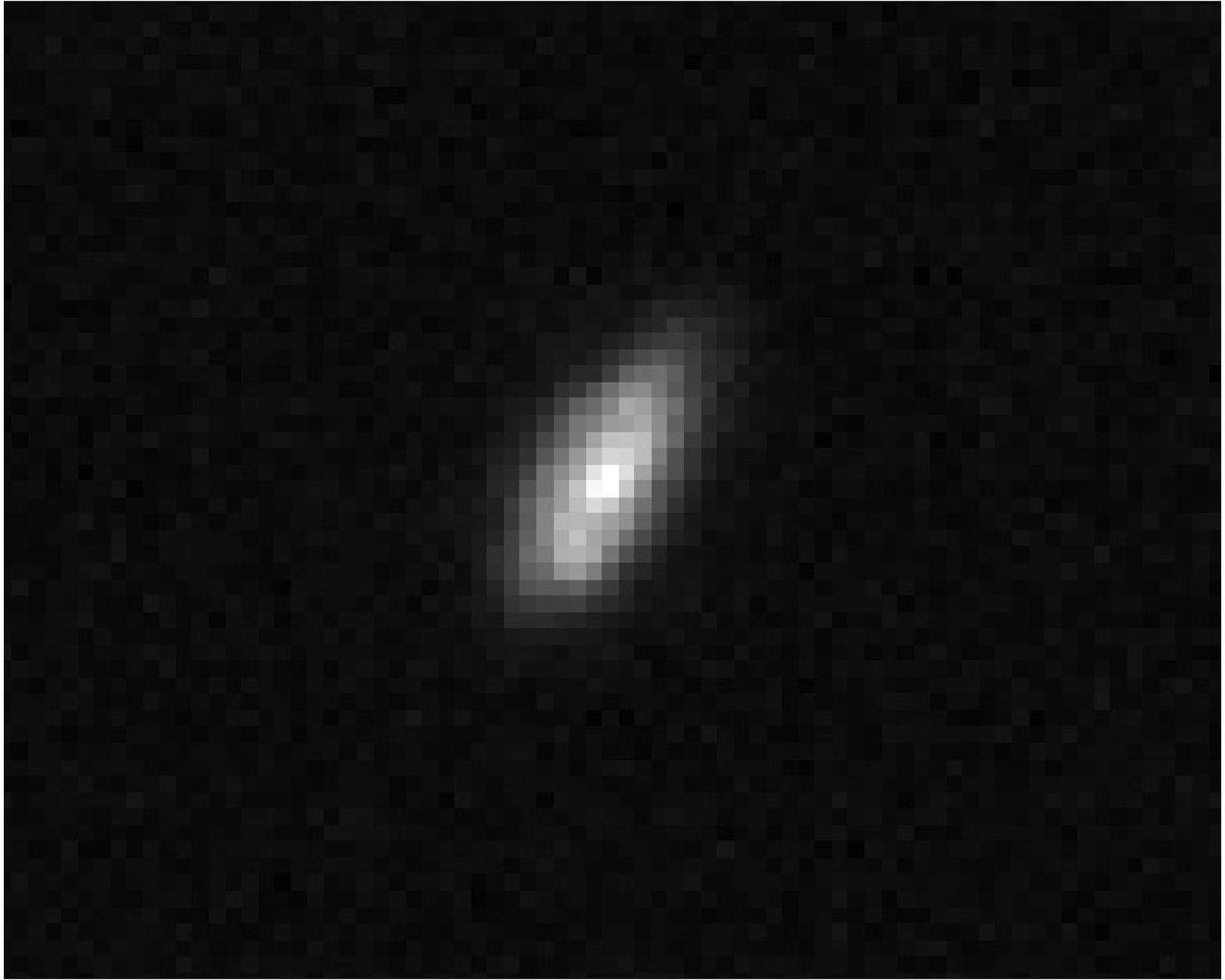}
\includegraphics[scale = .09]{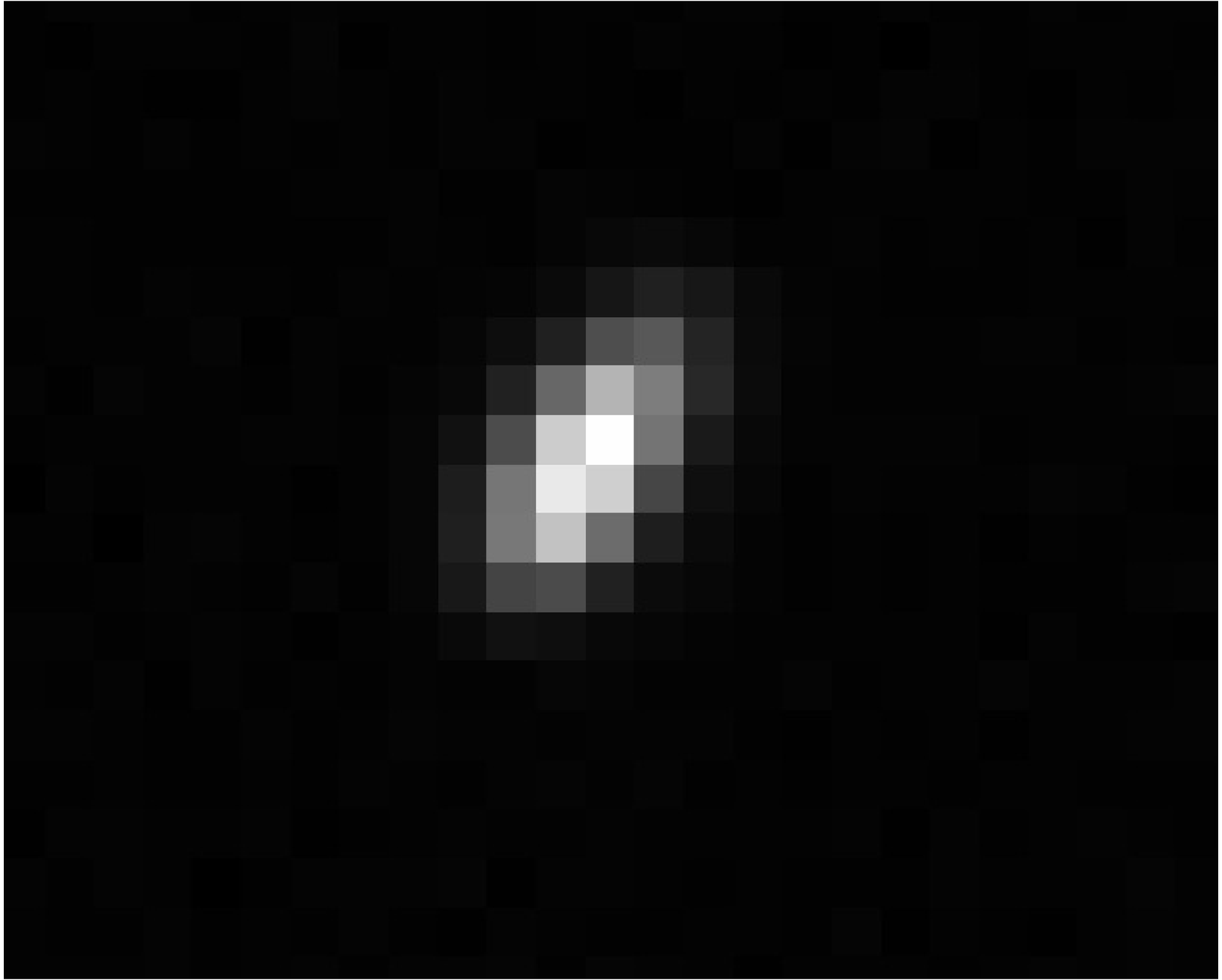}
\includegraphics[scale = .09]{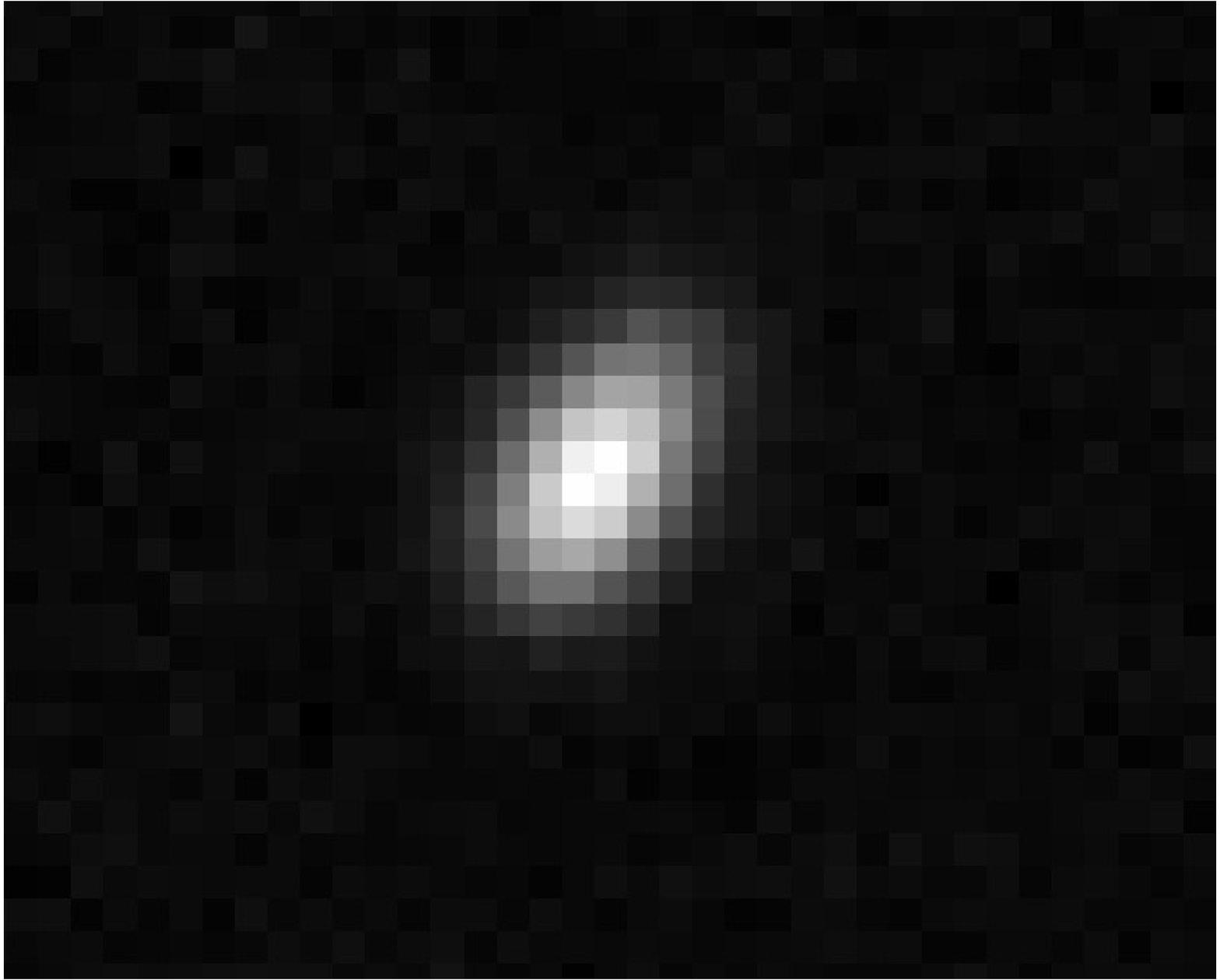}
\end{centering}
\caption{A galaxy observed with Euclid in the RIZ band (left), with
  Euclid in the H band (centre), and with Subaru in the i band (right)
  .}
\label{example}
\end{figure}

For the purpose of this work, we simulate the observation of the same patch of the sky with different instruments. As an example of future space-based survey, we choose to consider the Euclid experiment \citep{euclid}, that consists of an optical imaging channel and a near-infrared photometry channel. In particular, we start from a $800 \arcsec \times 800 \arcsec$ image of a simulated sky observed with a large optical imaging band analogue to the RIZ of the Euclid experiment (see Fig. \ref{rizimage}). Simulations made with Skylens of Euclid RIZ images are currently being used to predict the galaxy number density available for shear measurements from the observations (Meneghetti et al., in prep.). We underline that this broad band is proposed by the Euclid experiment to maximize the galaxy density used in cosmic shear measurements, but it is practically useless in photometric redshift determination, so the results of this work do not depend strongly on this choice. We then simulate the near-infrared coverage of the same sky, in particular considering three bands (Y J H). Characteristics of the filters and exposure times are taken from the Euclid proposal \citep{euclid}. Finally, we simulate the observation of the same sky with different ground-based instruments and different observing conditions. In the following, we explain how we analyse these images to extract the redshift estimation. In Fig. \ref{example} we show as an example the appearance of the same galaxy in different observing bands and conditions.

\subsection{Multiband photometry}\label{multiphot}

Our intent is to submit the images to software pipelines usually adopted for the analysis of real observations, to thoroughly simulate the extraction of a redshift estimate from images taken with different instruments. A fundamental step to get a good photometric redshift estimate is to correctly measure the colours from multi-band observations of the same galaxy. As the images have different PSFs and different levels of noise, measuring naively the magnitude in each band in a matched aperture is not appropriate. Instead, we use the Colorpro software, presented by \citet{coe}, that measures all the magnitudes in the same aperture, and then corrects them for the different PSFs. This software has been used in many experiments, including some that combine space-based and ground-based data, as in our case \citep[e.g.][]{colorpro1,colorpro2}.

In particular, we use the image of the galaxy in the RIZ band as the detection image, where the galaxy is detected and its aperture defined. To catch possible optical dropouts, we detect galaxies also in the H-band image, and we merge the two catalogues. Then the images in all the bands are remapped to the higher resolution of the RIZ image, and SExtractor MAG\_ISO magnitudes are measured from these images, considering the apertures previuosly defined. We underline that all the apertures we use for the photometry are drawn from the space-based RIZ image, to take advantage of the better PSF, and are tipically small. In fact, the distribution of apertures as a function of number of pixels has 95 as median, that corresponds to a radius $\sim 0.5 \arcsec$ in case of a circular shape. To estimate the amount of flux lost in each band due to the worse seeing, the RIZ image is degraded to the PSF of each band, and its new MAG\_ISO measured. The difference between this magnitude and the MAG\_AUTO measured on the original RIZ frame is the desired correction. This procedure is rigorously correct only if the surface brightness distribution is the same in all bands. This is true by construction in our simulations, because the object morphology is not dependent on wavelength. All magnitudes stated in this paper are AB magnitudes.

\begin{figure}
\begin{centering}
\includegraphics[scale = .7]{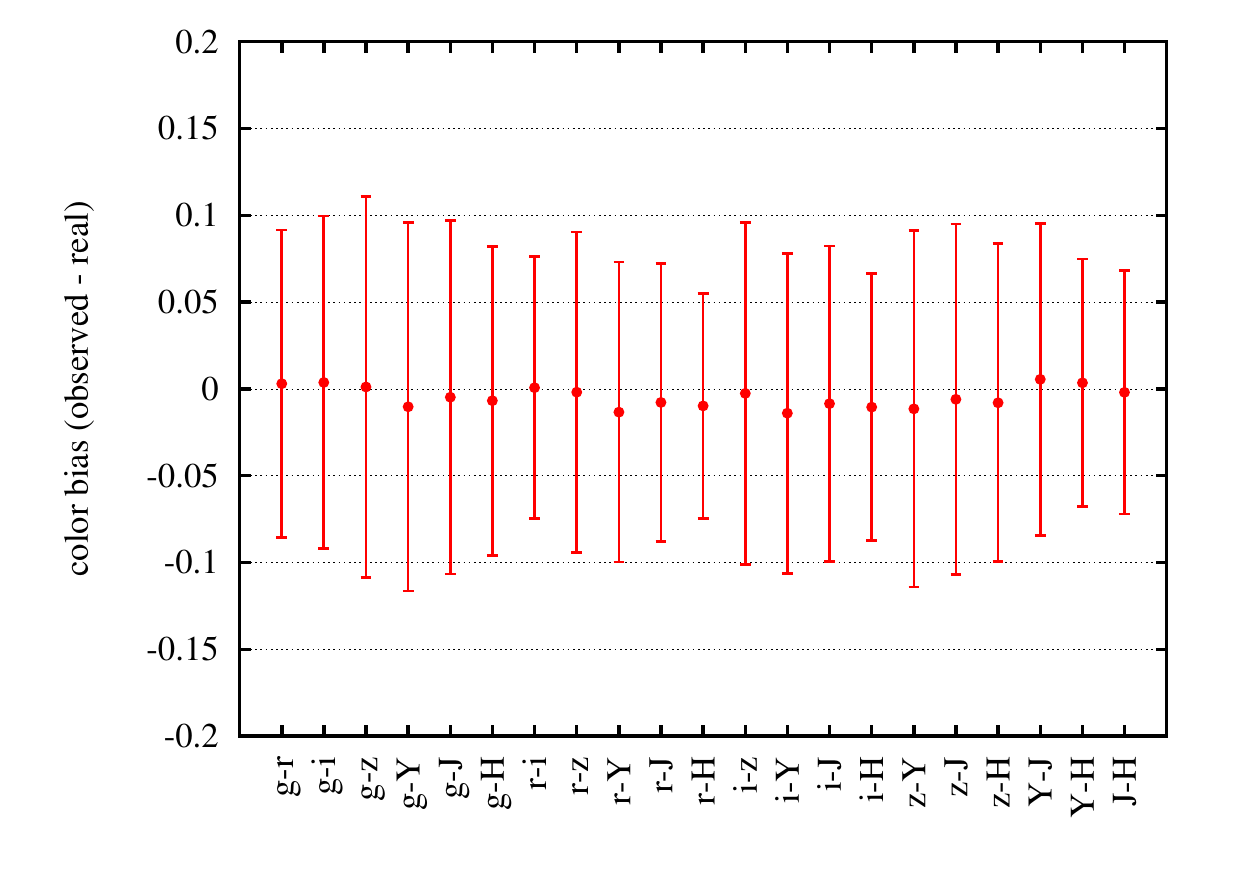}
\end{centering}
\caption{Precision in the colour measurement of our simulated images
  (Euclid + LSST) by Colorpro. For every couple of bands we show the
  mean of the colour bias over the galaxy sample, while the error bars
  represent the standard deviation of the measurement. This test has been performed observing one galaxy at a time, to avoid pollution from neighboring galaxies.}
\label{colors}
\end{figure}

\begin{figure}
\begin{centering}
\includegraphics[scale = .7]{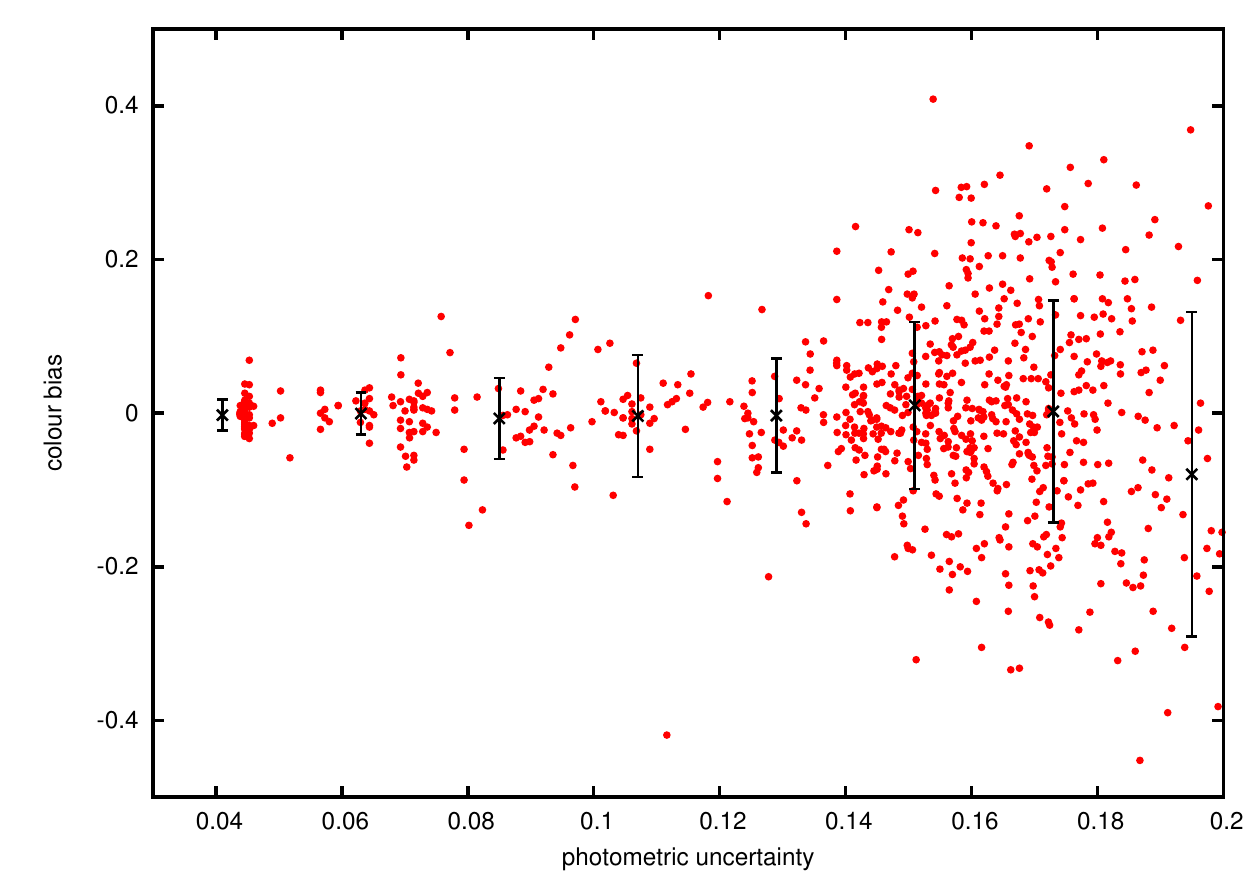}
\end{centering}
\caption{Precision in the colour measurement (LSST $g$ band - Euclid J band) of our simulated images
  by Colorpro. For each object we show the colour bias as a function of the photometric uncertainty in the colour measurement. The black bars show the mean and the standard deviation of the measurement in bins of photometric uncertainty.}
\label{colors_new}
\end{figure}

An example of the precision in the colour measurements obtained
applying Colorpro to our simulations is shown in Fig. \ref{colors},
where we plot the mean offset of measured colours with respect to the
input values. We considered combinations between the four space-based bands (RIZ Y J H) and the ground-based bands that will be used in the photometric redshift estimation ($griz$). For the purpose of this plot, we simulated the
observation of a small field around one galaxy at a time, to avoid
blending and pollution and thus check the robustness of the photometry
estimation process in an ideal case. We see that in this conditions
the mean bias is negligible and the standard deviation is below 0.1 magnitudes. We then focus on a single colour ($g$-J) and plot in Fig. \ref{colors_new}  the error in the colour measurement as a function of the photometric uncertainty for each object in the sample. We see that most of the objects have very small values of bias and that the statistical standard deviation is in good accordance with the uncertainty quoted by the program.  This means that the implemented method is able to compensate for the different PSFs and pixel sizes of the images.

\subsection{Redshift estimation}

With the procedure described above, we have obtained realistic
multi-band catalogues, with luminosity measurements in the 4 space-based
bands (RIZ,Y,J,H) and in the proper ground-based optical bands.  Now we
use these catalogues in the proper estimation of photometric redshifts for the galaxies. As before, we perform this
task applying a method that is developed for and usually applied to
real data. Different techniques have been proposed to extract the
redshift estimate of a galaxy from its multi-band photometry, such as
template fitting \citep[e.g.][]{bolzonella,bpz,feldmann} and
training-set methods \citep[e.g.][]{annz,way}. They are probed to obtain similar results \citep{abdalla2}, with the
former being more flexible, because they do not need to be trained with a complete spectroscopic subsample.
For this work we use the BPZ software \citep{bpz}, the implementation of a Bayesian template-fitting method, that has been applied to real data in many works \citep[e.g.][]{bpz1,bpz2}. It requires as input a set of galaxy template SEDs, from which the program calculates the colours when observed through the considered filters at different redshifts. The probability of a galaxy with colours $C$ of being at redshift $z$ is then evaluated applying
\begin{equation}\label{pz}
p(z|C,m_0) \propto \sum_T p(z,T|m_0) p(C|z,T)
\end{equation}
where $m_0$ is the magnitude used to compute the prior and $T$ is the index that runs over the templates. In particular, $p(z,T|m_0)$ is the prior given by the $i$ band distribution, as observed in the Hubble Deep Field North, while $p(C|z,T)$ is the likelihood of the colours given the template and the redshift, calculated via a $\chi^2$ approach,
\begin{equation}\label{like}
- \log {p(C|z,T)} \, \propto \, \chi^2(z,T,a) = \displaystyle \sum_\alpha  \frac{(f_\alpha - af_{T\alpha})^2}{\sigma^2_{f_\alpha}},
\end{equation}
where $f_{T\alpha}$, $f_\alpha$ and $\sigma_{f_\alpha}$ are the theoretical flux of the template, the observed flux and the noise in the band $\alpha$, respectively, and $a$ is the normalisation of the template. The BPZ software has been proven to perform competetitevely with other methods in \citet{abdalla2}.
Further details and a little modification of the original algorithm can be found in the Appendix.

In our case, we know the SEDs of our simulated galaxies, so we give the \textquoteleft right' template SEDs to the program. This set is formed by the six templates used by \citet{bpz}, plus the two added by \citet{coe} to fit the bluest galaxies of the HUDF sample. This is an ideal case, as in photo-$z$ template-fitting algorithms the correct choice of the template set is one of the critical steps \citep[][]{yee, swire}. The input magnitude that BPZ uses to compute the prior is the one measured in the $i$ band by the appropriate ground-based instrument.

\subsection{Catalogue cleaning}

\begin{figure}
\begin{centering}
\includegraphics[scale = .7]{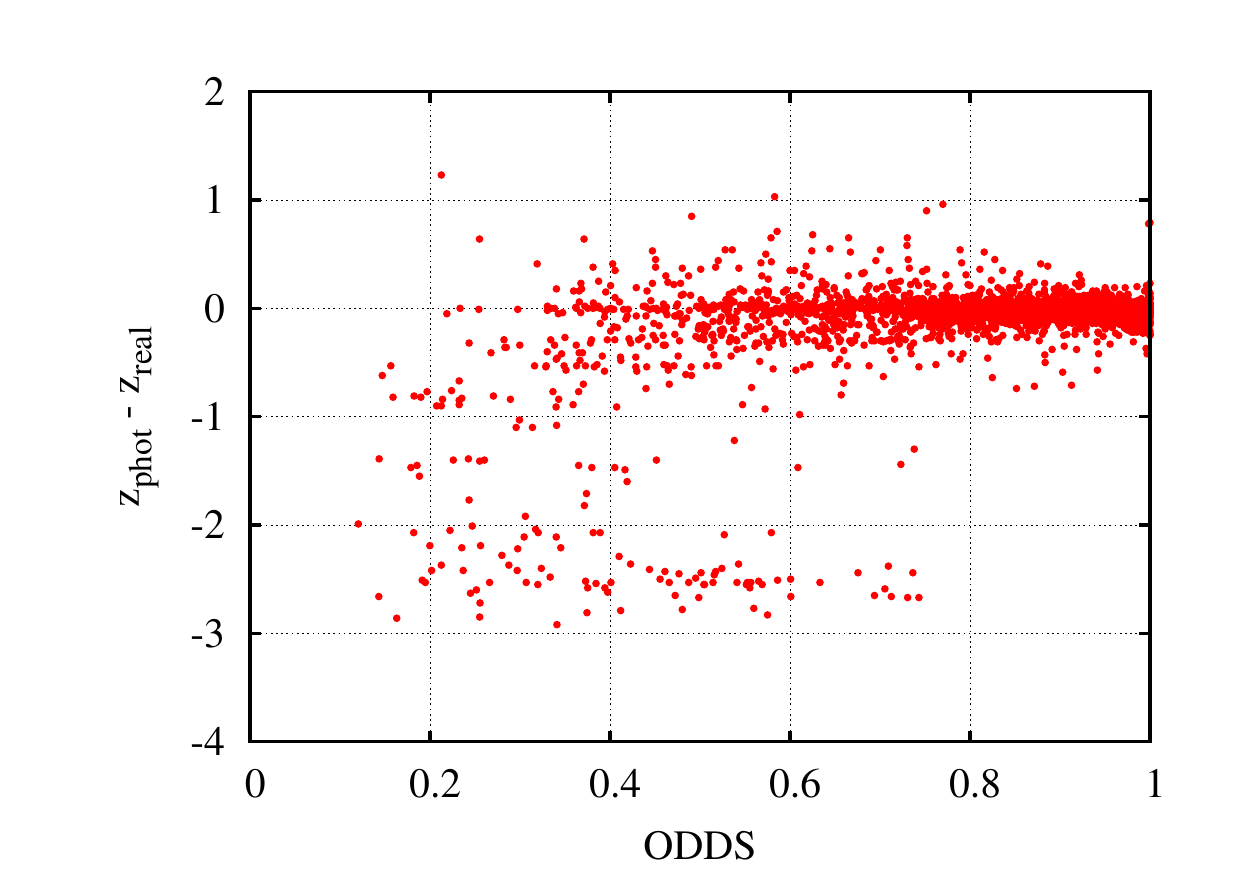}
\end{centering}
\caption{Example of the power of the ODDS parameter in the
  identification of catastrophic failures. The error of the
  photometric redshift estimate is plotted against the ODDS value
  computed by BPZ. Data taken from the 100 sec observation with
  the ideal telescope, $u$ band excluded (See Section 3.)}
\label{odds-deltaz}
\end{figure}

Photometric redshift estimation suffers from colours degeneracy for galaxies at distinct redshifts. Although the Bayesian prior implemented in Eq. \ref{pz} reduces this problem, e.g. avoiding that very brilliant galaxies are given high redshift estimates, it does not solve it completely. When two templates at different redshifts have very similar theoretical colours, photon noise will sometimes move the observed magnitudes of a galaxy towards the wrong template, and the redshift extracted from the algorithm will be
catastrophically wrong. Because of this, to match the strong requirements on the precision in the photometric redshifts for cosmic shear surveys, we need to clean our catalogue from galaxies that are likely to have a wrong redshift estimate. A way to do so is looking at the full redshift probability distribution of the galaxy, instead of considering only the most probable value. Galaxies that have colours in regions of degeneracy, i.e. that lie near the theoretical colours of templates at different redshifts, will have a multi-peaked probability distribution. Galaxies that have a controversial photometry, for example because are polluted by brilliant neighbours, will not show a single strong peak, but a diffuse distribution with much of the probability outside the vicinity of the most probable redshift. Thus, measuring the integral of the redshift probability distribution of the galaxy in the proximity of the most prominent peak will give an indication of how certain is the redshift estimation. This can be done via the ODDS parameter calculated by BPZ. It is defined as
\begin{equation}
$ODDS$ \, =  \displaystyle \int_ {z^\prime - 2\Delta_z} ^ {z^\prime + 2\Delta_z} p(z),
\end{equation}
where $z^\prime$ is the most probable redshift value, and $\Delta_z$ is
a free parameter, set by default in BPZ to $0.067(1+z^\prime)$. ODDS is thus a
measure of how much the distribution is concentrated around the
principal peak. This is indeed a powerful way to assess the goodness of the
estimate, as can be seen in Fig. \ref{odds-deltaz}, where we plot for each galaxy the error in the redshift estimate against the ODDS parameter calculated from its $p(z)$. The fraction of
galaxies with a catastrophic redshift estimate decreases rapidly as
ODDS increases.

If ODDS $>$ 0.95, it means that the
distribution is more peaked than a Gaussian with a width equal to
$\Delta_z$. This may seem a reasonable cut for the catalogue, and has
been used in some works \citep{coe, hilde}, while \citet{erben}
preferred a looser limit of 0.90. In the example of
Fig. \ref{odds-deltaz}, we see that catastrophic failures become rarer for galaxies with an increasing ODDS value, and disappear for galaxies with ODDS $>$ 0.80. In general, we need to trade off the amount of galaxies to be kept in the catalogue with the mean accuracy of the redshift estimation.
In the following, we will try to optimise the selection, by looking
at how the standard deviation of the redshift measurements varies as a function
of the ODDS value used as a minimum threshold. An optimised selection
is crucial in cosmic shear observations, as it is important to
consider as more galaxies as possible to reduce the noise induced by
the intrinsic shape of the galaxies.

\begin{figure}
\begin{centering}
\includegraphics[scale = .7]{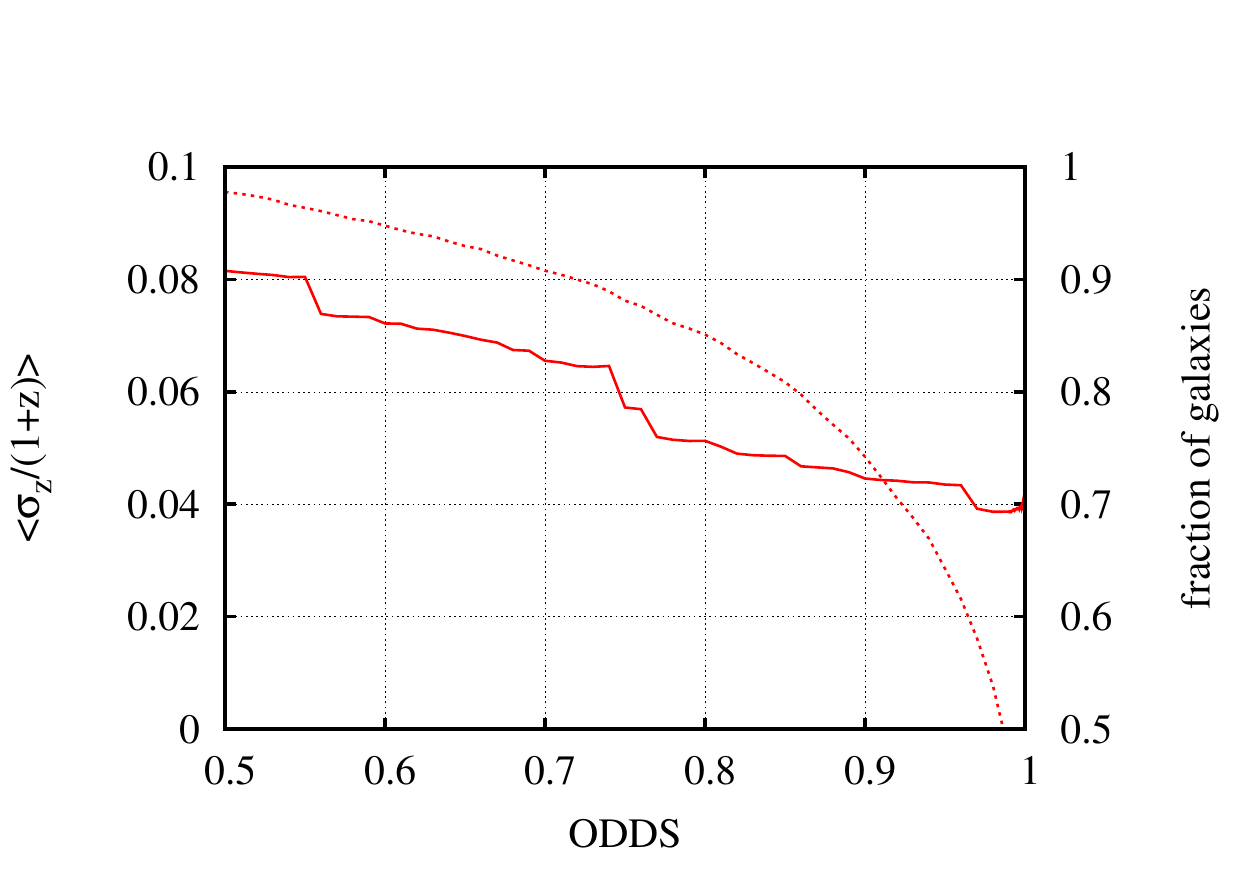}
\end{centering}
\caption{Normalised standard deviation (solid line) and fraction of available
  galaxies (dashed line) as a function of the ODDS value used as the minimum threshold for
  the photo-$z$ catalogue. Data taken from the 100 sec observation with
  the ideal telescope, $u$ band included (See below.) In this case, we must cut the catalogue at ODDS $\simeq$ 0.80 and eliminate $\simeq$ 15 \% of the sample to get to the required precision, $\langle \frac{\sigma_z}{1+z} \rangle = 0.05$.}
\label{odds-thresh}
\end{figure}

In particular we will measure the precision of the redshift estimates calculating the normalised standard deviation
\begin{equation}
\displaystyle \big\langle \frac{\sigma_z}{1+z} \big\rangle = \displaystyle \frac 1 N \bigg[\sum_{i=1}^N \frac{(z_{i,real}- z_{i,phot})^2}{1+z_{i,phot}} \bigg]^{1/2} ,
\end{equation}
where the index $i$ runs over the $N$ galaxies of the sample in the redshift range $0.3 < z < 3$.
We fix the requirement on the precision in the redshift measurement to $\langle
\frac{\sigma_z}{1+z} \rangle < 0.05$, that is considered valid to extract cosmological information from cosmic shear experiments, such as Euclid \citep{euclid}. We will
measure how many galaxies we must eliminate from the sample via a
selection based on the ODDS parameter to fulfil this condition. As an
example, in Fig. \ref{odds-thresh} we plot the value of $\langle
\frac{\sigma_z}{1+z} \rangle$ as a function of the ODDS value used as
a lower threshold, for one of the simulated observations. In practice, we
continue to throw away galaxies from the sample according to their ODDS value until
the normalised standard deviation of photo-z estimates goes below 0.05. In this
procedure, we will unavoidably eliminate some of the galaxies with a good photo-z
estimate, and we will as well keep in the catalogue some outliers, as
the indication given by ODDS is obviously only valid in a statistical
sense. Nonetheless, the mean standard deviation drops almost monotonically increasing
the ODDS threshold, so in this way we are in the end able to extract a robust galaxy sample
from the data.

For every survey setup our main result will be the ratio of the
number of galaxies that are kept in the catalogue after the cut over
the total number of galaxies in the initial sample. We will take into account only galaxies with a
measured RIZ magnitude below the declared limit of the Euclid mission
(24.5).
Although with a Euclid-like setup we
are able to go slightly deeper than this limit in the optical imaging
observation, some tests we did including dimmer galaxies proved not to
improve our results, probably because they suffer from poor infrared
photometry, so we decided to concetrate on a RIZ-magnitude limited
sample.
To increase the statistic, we create 1000 random realisations of each catalog via bootstrap resampling, and report the mean of the measurements.

\section{Dependence on survey parameters}

In order to obtain general results that do not depend on the specific
instrument or survey, we first simulate some optical ground-based multi-band
observations taken with a fictional ideal instrument. This instrument
has a very large collecting area (diameter = 8.2 m), a pixel angular
size of 0.2 arcsec and does not suffer from any losses due to mirrors and optics. The considered filters are the standard SDSS $ugriz$.
The performances of this irrealistic setup are compared with the ones of real telescopes considered in
this paper in Tab. \ref{eff}. Given the exposure time in a certain band with the ideal telescope, it is easy to calculate what is the time needed to reach the same depth with another telescope, considering the efficiency and the area of the instruments.

\begin{table}
\centering
\begin{tabular}{l|cccc}
\hline
& ideal & CTIO & PAN-STARRS & LSST \\
\hline
Area ratio & 1. & 0.149 & 0.193 & 0.628 \\
u eff.& 1. & - & - & 0.533  \\
g eff. & 1. & 0.539 & 0.478 & 0.713 \\
r eff. & 1. & 0.615 & 0.679 & 0.786 \\
i eff. & 1. & 0.669 & 0.706 & 0.813 \\
z eff. & 1. & 0.484 & 0.316 & 0.360 \\
Pixel size & 0.2 & 0.27 & 0.3 & 0.2 \\
\hline
\end{tabular}
\caption{Comparison of the performaces of the considered telescopes
  and optics. Efficiencies of the ideal telescope are equal to 1 by definition.}
\label{eff}
\end{table}

We underline that all the results presented in the following are obtained using data from the 4 space-based bands (RIZ Y J H), plus the indicated ground-based data.

\subsection{Exposure time}

\begin{figure}
\begin{centering}
\includegraphics[scale = .7]{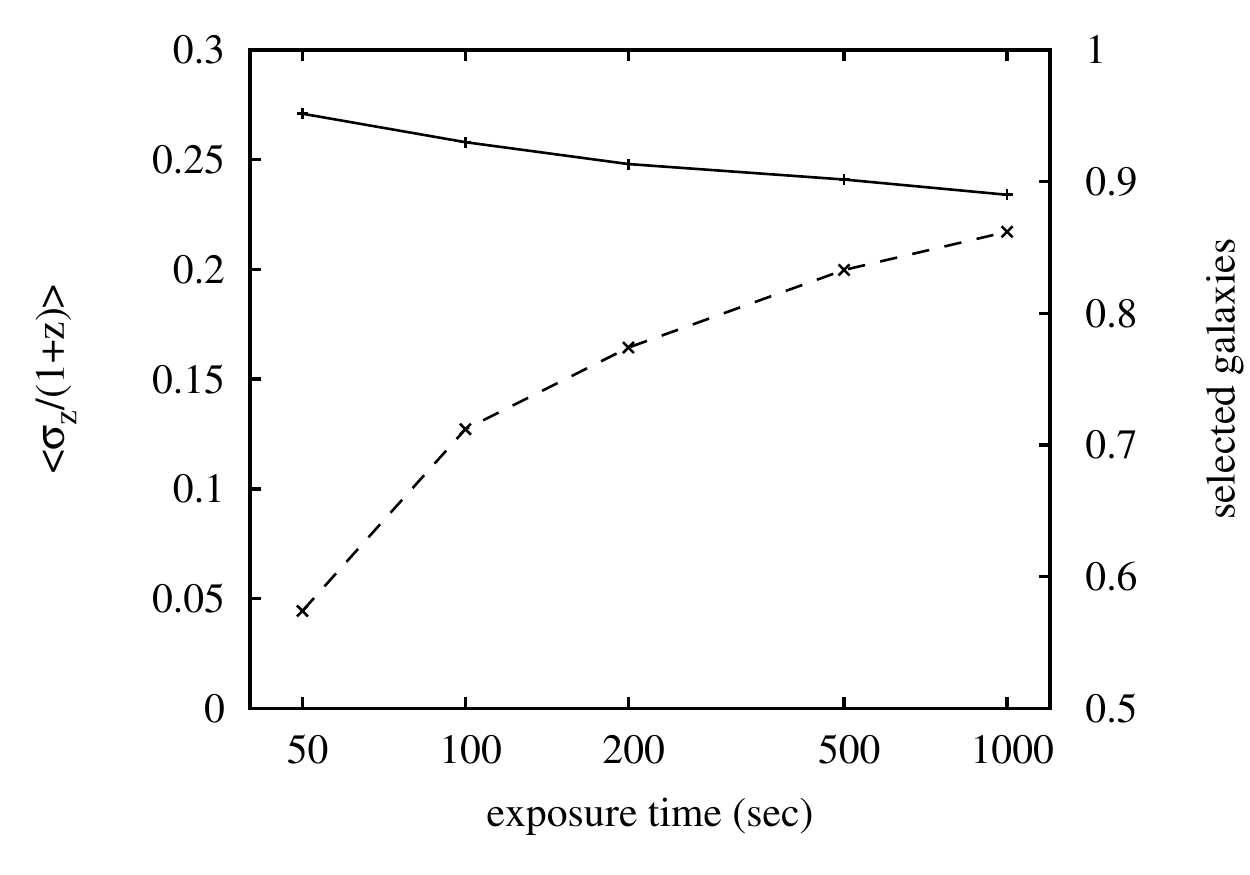}
\end{centering}
\caption{Quality of the photometric redshifts obtained with the $griz$
  bands of the ideal telescope as a function of the
  exposure time. The solid line represents $\langle \frac{\sigma_z}{1+z} \rangle$ as a function of the observing time, while the dashed line represents the fraction of galaxies with observed RIZ magnitude $<$ 24.5 that are available after a cut is performed to reduce the redshift standard deviation below 0.05. }
\label{tobs}
\end{figure}

The quality of the photometry, and thus of the photometric redshifts,
depends primarily on the depth of the survey. In this view, we
simulate a set of observations with a constant exposure time in each
$griz$ band. (The importance of the $u$ band will be studied separately in the
next Section.)
The results of the simulation are shown in
Fig. \ref{tobs}, where we show the standard deviation of the total sample (i.e.
formed by all the galaxies with RIZ magnitude $<$ 24.5, without any
selection based on ODDS), and the fraction of galaxies that are kept
in the catalogue after a selection analogue to Fig. \ref{odds-thresh}
is performed. We see that the standard deviation of the whole
sample does not improve significantly as the exposure time increases, but
there is a great enhancement of the number of available galaxies after the
cut is performed. This happens because the standard deviation of the total
sample depends mainly on catastrophic failures due to redshift-colours
degeneracies. These are only partially avoidable improving the quality
of the photometry, because they are an intrinsic drawback of the photometric redshift
estimation. Above this, we have a number of galaxies with poor
photometry due to bright companions or to blending with close dim
neighbours. In this case, the observed flux is the sum of two
different objects, and thus the observed SED will likely not
correspond to any of the two physical states. Luckily, the selection
via the ODDS parameter allows us to reject these galaxies. The
improvement in the fraction of available galaxies after the selection for longer exposures
is then given by the more precise photometry that we obtain. Reducing the errors on the measured magnitudes, the
non-pathological galaxies
(i.e. not exactly on a redshift-space degeneracy and without close
companions) will be assigned a redshift with a smaller
uncertainty. Thus they will have a more peaked redshift probability distribution and a bigger ODDS parameter. This will avoid them from being confused with catastrophic failures when the selection is performed. We note that with the
deepest exposure, 1000 sec, just the 14\% of the total sample of
galaxies brighter than 24.5 in the RIZ band must be excluded to
meet the requirements in the redshift precision.


\subsection{Observing bands}

\begin{figure}
\begin{centering}
\includegraphics[scale = .7]{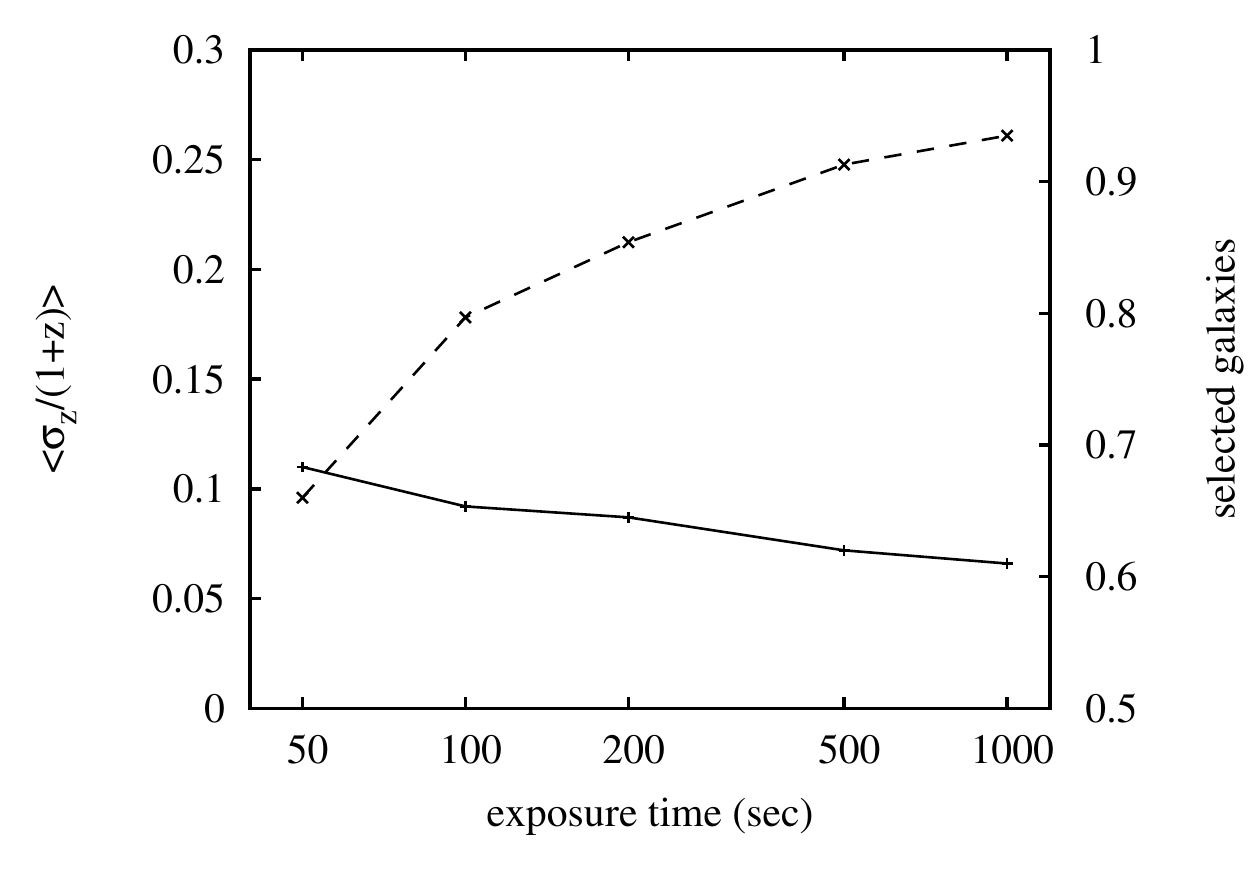}
\end{centering}
\caption{As Fig. \ref{tobs}, but including also the $u$ band data.}
\label{tobs_u}
\end{figure}

It is sometimes possible to extend the observed wavelength range using
the $u$ band. This band is more time consuming then the others due to
the increased absorbtion of the atmosphere with respect to the redder wavelengths
and to the lower quantum efficiency of the CCD detectors. It is then usually necessary to trade off the availability of data at
shorter wavelength with the loss of depth in the other bands. To test
the importance of $u$-band data, we run the BPZ algorithm on the same
images of the previous section, but with an additional $u$-band
observation of the same exposure time. The results are shown in Fig \ref{tobs_u}.

\begin{figure}
\begin{centering}
\includegraphics[scale = .7]{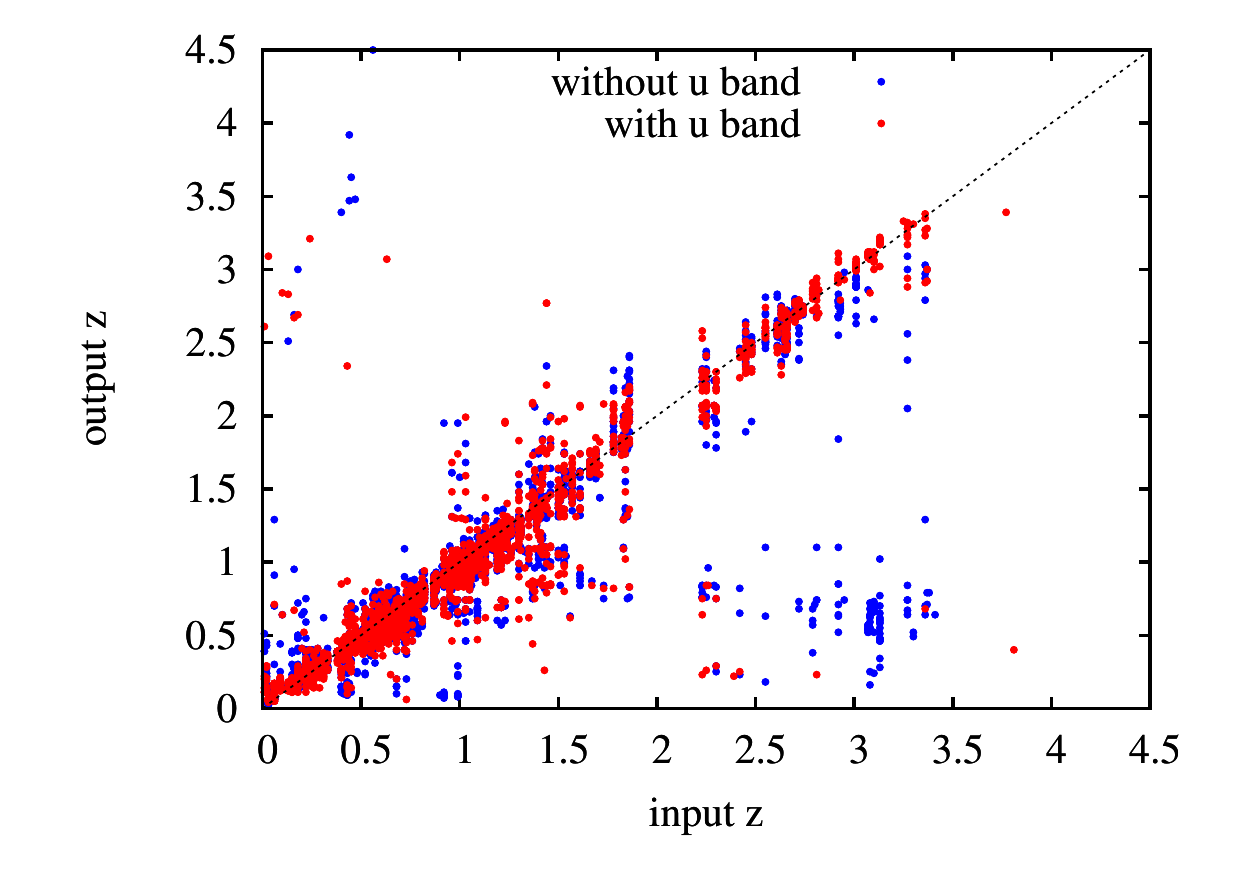}
\end{centering}
\caption{Estimated redshift against input redshift of the sample galaxies, considering the $u$ band
 (red points) or discarding it (blue points). Data are taken from the
  observations with an exposure time equal to 100 sec per filter. The alignment of some group of points along the $y$-axis is caused by the repetition of the same galaxy in different sky positions, due to the limited amount of galaxies in the initial HUDF catalogue.}
\label{nouy}
\end{figure}

We see that even the shorter exposure that includes the $u$ band (50 sec, corresponding to a 10-$\sigma$ depth equal to 23.5)
gives a total redshift standard deviation lower than with the deepest exposure with the $griz$ bands only. This
result is not surprising as the power of the $u$ band in photometric redshift estimation is widely recognised in the literature \citep{abdalla,dahlen} . In
particular, it helps in identifiying galaxies at low redshift (z $\lesssim$
0.5) thanks to the ultraviolet continuum, and the ones at z $\sim$ 3 when the
Lyman break enters the blue side of the filter. The realisation of
this can be seen in Fig. \ref{nouy}, where we plot the output redshift
against the input one, both with and without using the $u$-band
data. The reduction of catastrophic failures especially for galaxies
at high redshift is very evident.

On the other hand, when considering the number of galaxies after the selection, we see that the results are closer to the ones with the $griz$ bands only, thanks to the fact that many catastrohpic failures are in that case pointed out and eliminated. The difference between the two setups is around 10 \% of the total number of galaxies for the whole range of exposure times tested. With the longest exposures, the $ugriz$ configuration permits to use around 94 \% of the sample.

\subsection{Seeing}

\begin{figure}
\begin{centering}
\includegraphics[scale = .7]{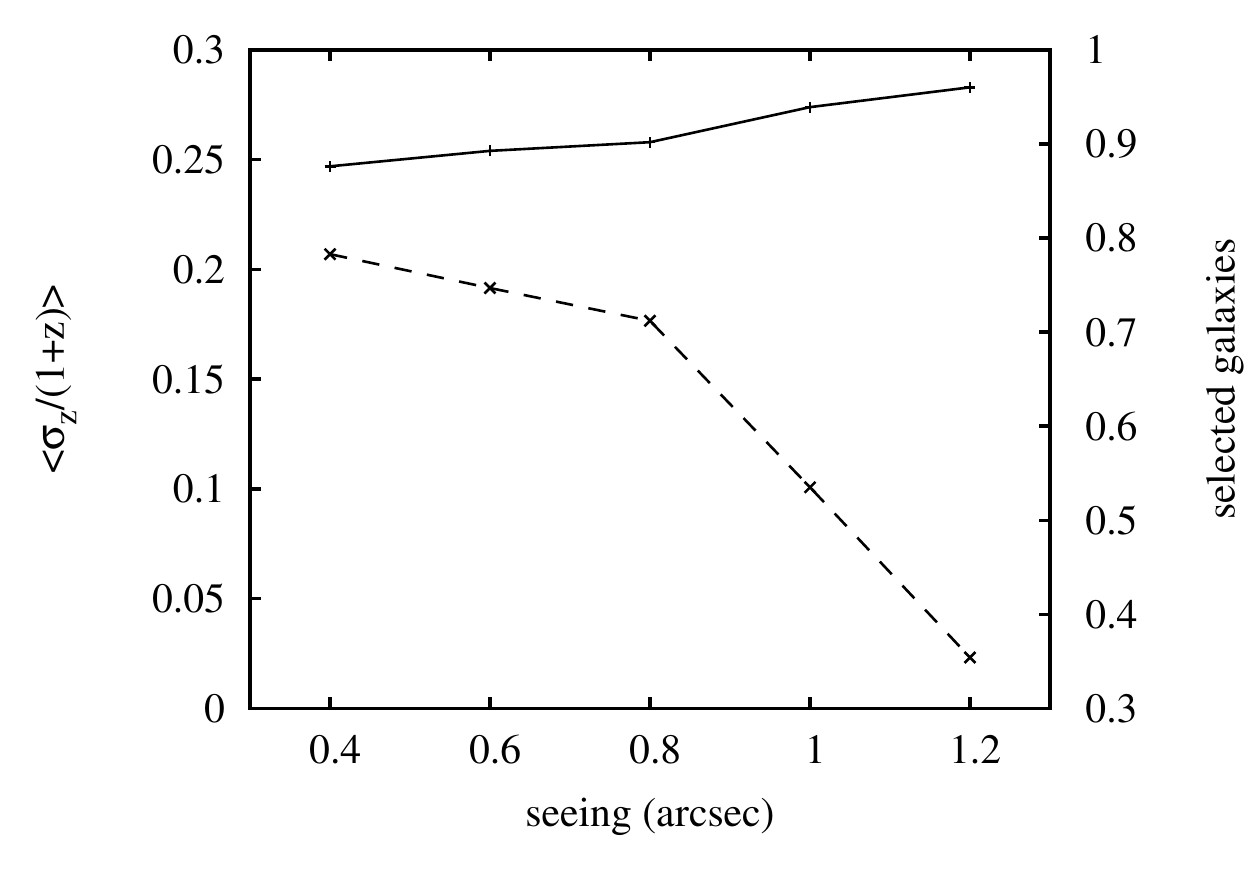}
\end{centering}
\caption{Quality of the photometric redshifts obtained with the $griz$
  bands of the ideal telescope as a function of seeing. The solid line represents $\langle \frac{\sigma_z}{1+z}
  \rangle$ as a function of the seeing, while the dashed line
  represents the fraction of galaxies with observed RIZ magnitude $<$ 24.5 that are available after a cut is performed to reduce the redshift standard deviation below 0.05. }
\label{see}
\end{figure}

A problem that influences the quality of the photometry in
ground-based observations is the blurring of the images that occurs as
the light travels through the atmosphere. This is often quantified by
the seeing, i.e. the mean FWHM of the observed image of an intrinsic point-like
source as a star. The method used to perform these simulations allows
us to study the influence of this parameter on the precision of the
photometric redshifts. In particular, we simulated five different
observations of 100 sec per filter with the ideal telescope, with a seeing varying from
0.4 to 1.2. Results are shown in Fig. \ref{see}.

As in the case of the observing time, the change in the precision of
the whole sample is not significant, while much more evident is the
improvement in the number of available galaxies after the cut is
performed. In fact, the number of galaxies with a certain redshift estimate can
change by a factor of two depending on the quality of the atmosphere of
the telescope site. This of course depends on the leakage of light
between different galaxies that occurs if the seeing is greater than
the angular distance between them. This creates an observed SED without a corresponding galaxy template and thus with a poor redshift estimate.

In reality, many astronomical images are the sum
of exposures taken in different moments and with
different observing conditions, i.e. without a single seeing value
that can be attributed to the whole image, as in our idealised work. Nonetheless, our results are a strong
indication that good observing sites are required to perform the high-precision photometry needed for redshift estimation of deep and vast galaxy catalogues.

\section{Planned surveys}


Different projects are considered as possible ground-based
counterparts of space-based weak lensing observations. We underline that any space-based mission that will observe both
the northern and southern sky will need collaboration from
different projects covering different regions of the sky. In some sense,
some of the projects we study will not be competitors against
each other in practice, because any space-based project will need a
patchwork of collaborations around the globe.

In this work we consider three ground-based surveys: the Dark Energy Survey (DES), Pan-Starrs and the Large Synoptic Survey Telescope (LSST). The first two are officially supporting
the Euclid mission, while the third is a case we chose in order to
explore the potential of observations that consider the $u$ band,
following the results of the previous Section. In Table \ref{obsTime}
we show the exposure times and the depths of the three surveys,
calculated with our simulator.

\begin{itemize}
 \item  DES is a project that will begin in late 2012 with the Blanco 4-meter telescope at Cerro Tololo
Inter-American Observatory, on which a new camera named Decam will be
mounted at the scope \citep{des}. It will observe 5000 square degrees in the South
Galactic Gap for 525 nights in 5 years.

 \item  Pan-Starrs is a new telescope,
consisting of four mirrors, one of which is already working at Mount
Haleakaka in Hawaii \citep{panstarrs}. It will observe the 30000 square degrees
visible from the site. The exposure times of Table \ref{obsTime} are calculated from the ones
declared for the ongoing 3-years 3$\pi$ survey \citep{pan1}, multiplied by 10/3, to make
them realistic for a 10-year project.

 \item  LSST is a telescope that will
start observing in 2017 from Cerro Pachon in Chile \citep{lsst}. It will
observe the southern half of the sky. The observing
times are the ones declared for the 10-year total observing project.
\end{itemize}

\begin{table}
\centering
\begin{tabular}{l|cc|cc|cc}
\hline
Band & DES & DES & PAN-ST. & PAN-ST. & LSST & LSST \\
 & time & depth & time & depth & time & depth\\
\hline
u & - & - & - & - & 1800 & 25.7\\
g & 500 & 24.3 & 2400 & 25.6 & 2400 & 26.6\\
r & 500 & 24.0 & 1200 & 24.9 & 5400 & 26.6\\
i & 900 & 24.1 & 1200 & 24.5 & 5400 & 26.2\\
z & 2100 & 24.0 & 1200 & 24.1 & 4920 & 25.7\\
\hline
\end{tabular}
\caption{Exposure time and 10$\sigma$ point-source sensitivities for
  the simulated surveys. Although some of the
experiments foresee to observe also in the $y$ band, we did not
consider it because its wavelength range is well covered by the space Y
band.}
\label{obsTime}
\end{table}

\begin{figure}
\begin{centering}
\includegraphics[scale = .7]{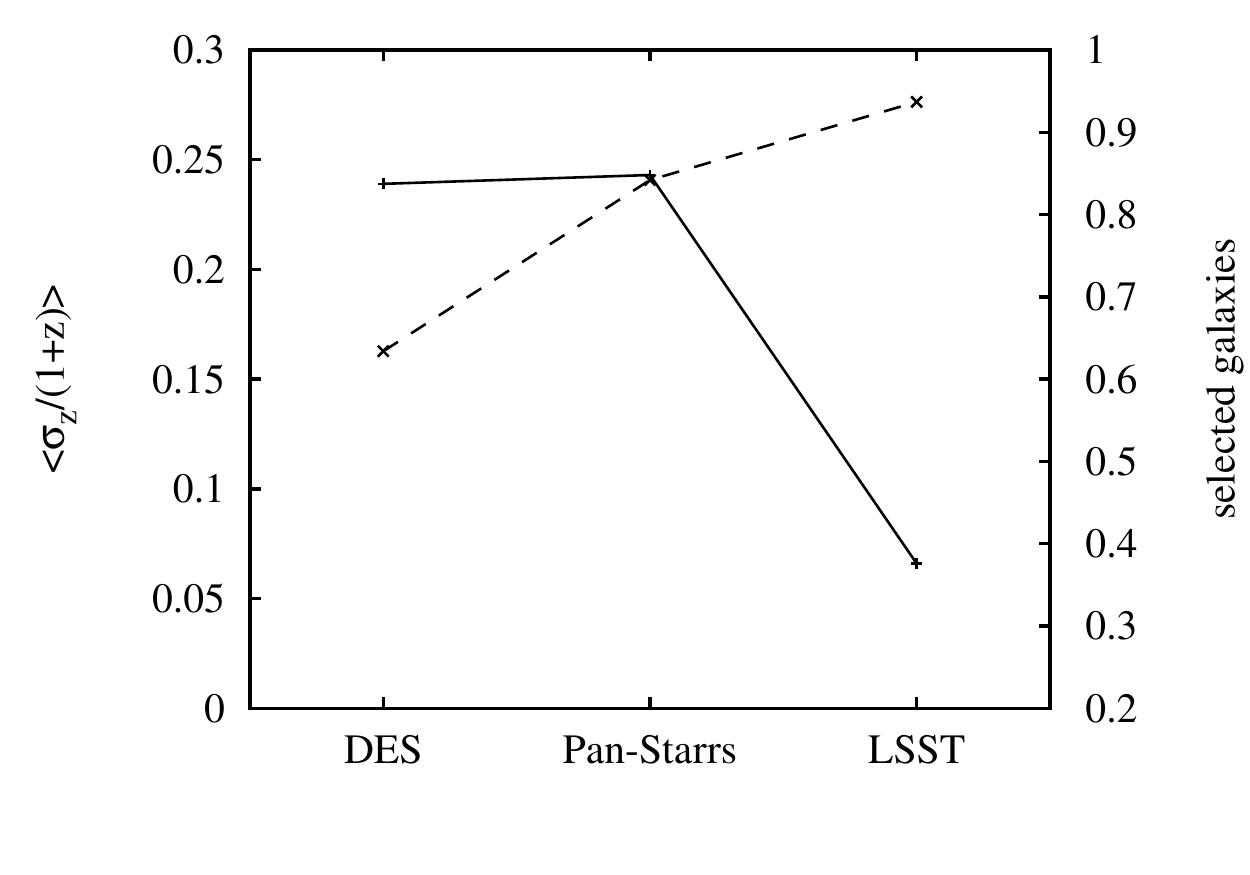}
\end{centering}
\caption{Quality of the photometric redshifts obtained with the
  different ground-based surveys. The solid line represents $\langle \frac{\sigma_z}{1+z}
  \rangle$, while the dashed line
  represents the fraction of galaxies with observed RIZ magnitude $<$ 24.5 that are available after a cut is
performed to reduce the redshift standard deviation below 0.05. }
\label{sur-sel}
\end{figure}

For each project, we perform the simulations as described in Section 2, keeping the
seeing fixed at 0.8 arcsec. This is a reasonable value, that can be
obtained in all three sites \citep{des_see, pan_see, lsst_see}. The results are given in Fig. \ref{sur-sel}.

\begin{figure}
\begin{centering}
\includegraphics[scale = .7]{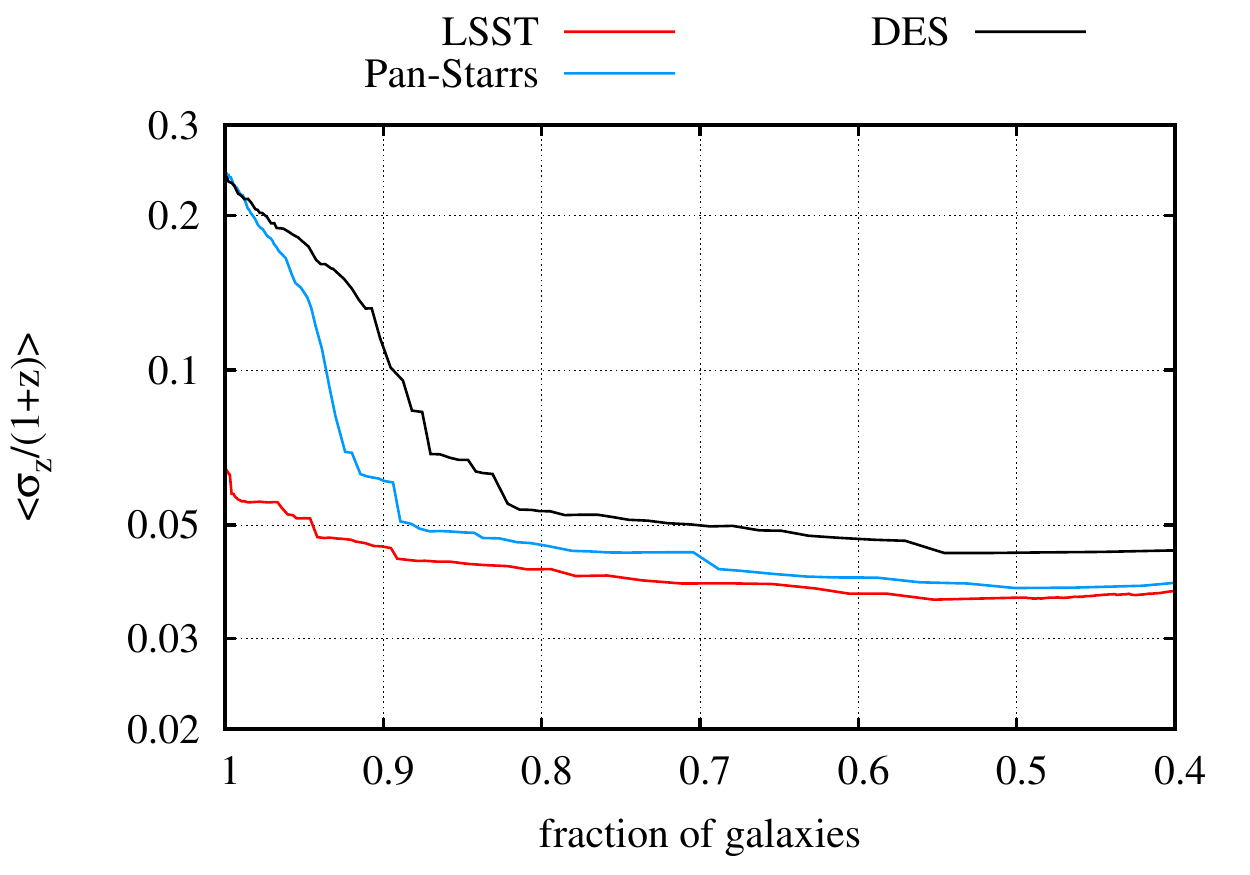}
\end{centering}
\caption{Redshift standard deviation as a function of the fraction of galaxies kept in the sample via a ODDS selection for the three simulated surveys. }
\label{odds_sel}
\end{figure}

As expected from the simulations with the ideal telescope, there is a big difference between the two $griz$ instruments and LSST, when concerning the standard deviation in the redshift measurement of the whole sample. On the other hand, the number of galaxies available after the selection differs a lot also between the two $griz$ instruments. The longer exposure times of Pan-Starrs make its results almost the best achievable for a $griz$ survey, with more than 80 \% of the galaxies retained after the selection. The LSST-like survey is the only one that goes beyond 90 \%, as could be guessed by the simulations with the ideal telescope.
A detailed plot of the redshift standard deviation as a function of the fraction of galaxies kept in the sample can be seen in
Fig. \ref{odds_sel}. From this figure it is possible to evaluate a possible trade-off between redshift precision and galaxy fraction for the different instruments. We see that, even with the most severe selection, the redshift standard deviation remains above 0.03 in all three cases. This is due to very unfortunate overlapping of galaxies, that were detected as single objects by our routines. The redshift deduced by the combined photometry of the two galaxies is then compared to the one of the most brilliant object in the RIZ band, that is sometimes totally different. This kind of problem depends very mildly on the number of bands and on the quality of the photometry and has been only partially solved running a detection on the H-band image, in addition to the one on the RIZ image (See Sect. \ref{multiphot}).

The trade-off between redshift precision and galaxy fraction is also strongly connected to the effectiveness of the procedure of removing galaxies that are likely to have a wrong redshift. We used a selection via ODDS, while other authors preferred using the 1-$\sigma$ error quoted by the photometric redshift code \citep{abdalla}, or investigating the presende of secondary peaks \citep{bordoloi}. We tried these other methods, but the results were in general poorer that the ones we obtained with ODDS. If we were able to make an ideal selection, knowing perfectly what are the most wrong redshifts in the sample, we would need to remove $\sim 1\%$ of the galaxies with the LSST data, and $\sim 5\%$ with the DES and PanStarrs ones, to get to the required precision.

\section{Comparison with previous studies}

\begin{figure}
\begin{centering}
\includegraphics[scale = .7]{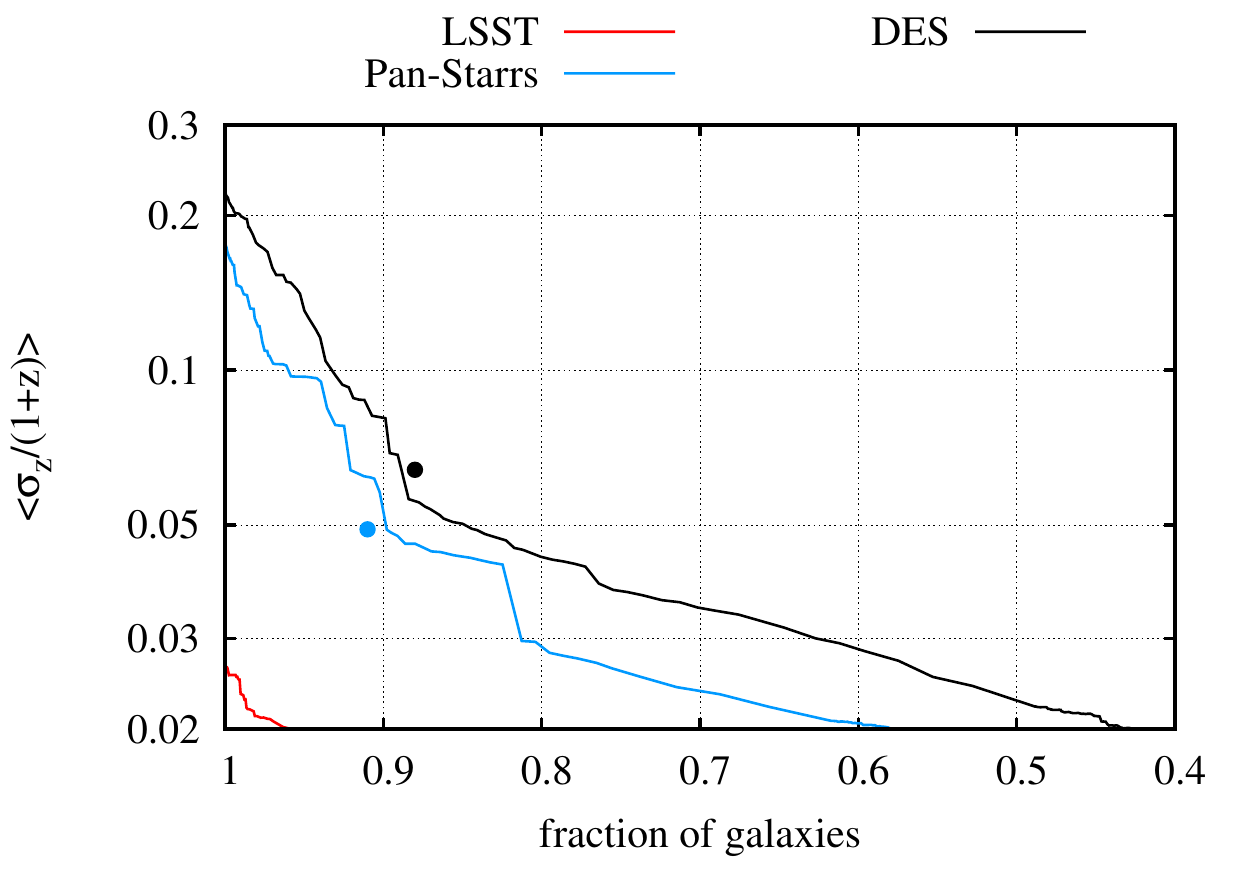}
\end{centering}
\caption{As Fig. \ref{odds_sel}, but from mock catalogues obtained without going
  through image simulation. The points represent the results obtained
  by \citet{bordoloi}.}
\label{odds_sel_fake}
\end{figure}

Our simulations are of course a simplified version of what is done to
measure redshifts from multi-band real images, nonetheless they are
more realistic than the simple creation of dirty catalogues. Thus, is
not surprising that our results are in general worse than what is
obtained by previous works done with that method
\citep{abdalla,bordoloi}. For example, \citet{bordoloi} find, for
a DES-like survey, a standard deviation of the selected sample equal
to 0.064 (so in fact higher than our threshold) after removing 12 \%
of the galaxies. As can be seen in Fig. \ref{odds_sel}, we instead would need to eliminate $\sim$ 15 \% of the
sample to get to the same precision. To
better understand this difference, we made some tests using simulated
catalogues created without going through image simulation. We
calculated magnitudes in the necessary bands from the theoretical SEDs
of the galaxies in our sample, and we cut the resulting catalogue at RIZ $<$ 24.5. Then,
we added Gaussian noise according to the observing depth for the
Euclid-like survey and for the three realistic ground-based
surveys. The resulting multi-band catalogues have been submitted to
BPZ and analysed in the same way as those obtained from Colorpro on
the simulated images. In Fig. \ref{odds_sel_fake} we show the results,
namely the redshift standard deviation as a function of the fraction of galaxies kept in the
catalogue. The results we obtain in this way are
significantly better than the ones obtained with the complete
simulation (Fig. \ref{odds_sel}), and are in close agreement with \citet{bordoloi}.

As can be seen from Figs. \ref{odds_sel} and \ref{odds_sel_fake}, the
improvement we obtain with the simpler set of simulations is relevant
especially in the cleaning procedure, that becomes more
effective. This may depend on the fact that in this case the errors quoted in the
photometry catalogue are by construction the real uncertainties on the
magnitudes, while the procedure of multiband
photometry when dealing with images can bring hidden errors, such as those due to the
matching of images with different quality. As the errors in the
photometry translate in the uncertainty in the redshift determination,
the value of the ODDS parameter is certainly more trustable in the
simpler case and cleaning the catalogue from possible catastrophic outliers is thus more straightforward.

The work of \citet{abdalla} differs in that they use the
neural-network code ANNz to compute the redshifts. This may explain
some differencies on the importance of different bands between this
work and theirs. In fact, they quote a $\sigma_z$ (note: without
dividing by $1+z$) of 0.307 for a DES survey,
0.274 for Pan-Starrs, 0.197 for LSST. We get 0.398 (0.359 with the
\textquoteleft simple' simulations) in the DES case, 0.400 (0.311) for
Pan-Starrs, 0.146 (0.131) for LSST. Thus, they perform slightly
better in the $griz$ cases, while we do better when the $u$-band is
available. This reflects also in the fact that they are able to get a
$\sigma_z$ = 0.113, below the LSST case, for an \textquoteleft ideal'
optical survey, while we do not obtain such a result, as shown in
Section 3. This is reasonably due to the completely different
techniques adopted. Although in general they get comparable results, it is possible that in this case we are showing strengths and
weaknesses on both sides. Of course a firm statement on this topic can
be done only submitting the same photometric catalogue to different
photo-$z$ algorithms.

\section{Summary and conclusions}

In this paper we investigated the quality of
photometric redshifts that can be obtained for the galaxies observed by future space-based missions targeted to measure cosmic shear. In order to create a realistic setup for our work, we chose to simulate a Euclid-like mission, with imaging in an optical band and multi-band photometry in the near infrared \citep{euclid}. Fulfilling the requirement on photometric redshift precision is a crucial step to reduce the error in cosmic shear measurement, and it is important as well in other scientific subjects that can be studied with the same data (e.g. galaxy clusters, galaxy evolution). To this end, the space-based survey will need the collaboration of ground-based multi-band observations to characterise the SED of the galaxies in the optical range. We studied this topic using complete
simulations, from realistic images of a patch of the sky, obtained with Skylens \citep{skylens},
through multi-band photometry, to the photometric redshift estimation
itself, performed with a template-fitting method \citep{bpz}. This allowed us to consider many sources of uncertainty in the
process of determining a galaxy photo-$z$. Our main aim was to determine
what fraction of the galaxy sample observed by a Euclid-like survey must be discarded to achieve $\langle
\frac{\sigma_z}{1+z} \rangle < 0.05$, a precision considered suitable for cosmic shear precision measurements.

We first studied the impact
of different parameters (depth, seeing, bands) on the quality of the
redshift estimation.  We found that:
\begin{itemize}
 \item The number of good photo-$z$s increases as expected with the
   exposure time, as more and more galaxies acquire a good photometry
   that enables to assign to them a certain redshift. The improvement
   is not so relevant if we instead consider the mean precision of the
   whole sample, as some pathological galaxies (affected by light of
   neighbouring objects, or by severe colour-redshift degeneracies) pollute the catalogue even with the deepest
   survey we simulated.
\item Seeing is potentially as important as depth, as blurring of
  the images is a great source of noise in the photometry, expecially for
  FWHM $>$ 1 arcsec. Increasing the seeing from 0.8 to 1.2 arcsec, we need to remove twice the number of galaxies to reach the same precision in photo-$z$ estimation.
\item Having $u$-band observations seems very important to improve the
  goodness of photo-$z$s. Even a relatively shallow survey with a 10-$\sigma$ magnitude limit equal to 23.5 has a great impact on the redshift standard deviation of the total sample. Ultraviolet light is particularly relevant for
  galaxies at redshift below 0.5 and above 2.5 and resolves a lot of
  magnitude-colour degeneracies between these two ranges. The impact is not so strong when dealing with the fraction of available galaxies: in this case the improvement is around 10 \% of the total sample in comparison with a corresponding $griz$ survey.
\end{itemize}
Then, we moved to some more specific tests, aimed to study the results
that would be obtained with some of the ground-based surveys that are
currently considered as possible Euclid counterparts. We focused on
three projects: DES, Pan-Starrs and LSST. DES and Pan-Starrs suffer
the lack of $u$-band observations, and to clear the catalogue from the
catastrophic failures and get to the required precison we need to
eliminate respectively $\sim$ 35 \% and $\sim 15$ \% of the galaxies of the sample. LSST performs much better
than the other two, and seems to be the best, even if the farthest in
time, counterpart to a space-based cosmic-shear mission.

With the aim of comparing our results with previous works, we made
also some simpler simulations, creating catalogues with known
photometric uncertainties without making mock observations. In this
way, as expected, the results improve significantly, especially in the
cleaning procedure. This means that the depth of the survey alone does not
include all the information about the noise of the observation. Other
uncertainties are added because of the proximity of galaxies, that
pollute each other's photometry, and to the necessity of matching images obtained
with different observing conditions and telescopes. A
possible solution to the problems caused by the proximity of galaxies would be to use some automatic
algorithm to eliminate a priori objects that appear to be very close
to each other or even completely blended (i.e., detected as a
single object). This points towards the need of even more realistic
simulations, where it will be possible to optimize the procedure of
object detection and multi-band photometry extraction from the images,
a topic we did not tackle in this work. For example \citet{kuijken}
proposed a more complex approach, that makes use of
shapelets decomposition to model both the source and the point-spread function.

We want to stress here that we used a photometric redshift code that has
performed well in other applications to real and simulated data, but
we did not select it as the best in the treatment of the specific data
set we applied it to, nor we optimised it in this view. Our sample
cleaning method could maybe be improved too, e.g. tuning the
integration limits in the ODDS calculation, or considering secondary
peaks, as done by \citet{bordoloi}, or the 1-$\sigma$ error quoted by the photometric redshift code, as done by \citet{abdalla}. In general, our results should not be considered as optimal forecasts for a Euclid-like
experiment, but more as a quantification of strengths and
challanges for this kind of mission. In particular, the difference
between results obtained with noisy catalogues (Fig. \ref{odds_sel_fake}) and with image
simulations (Fig. \ref{odds_sel}) hints at the importance of multi-band photometry
extraction and \textit{a-priori} object selection for this kind of measurement. Further testing and development
of the algorithms is needed to understand what is the optimal way to
extract photometric redshifts from this kind of observations and
to optimise the methods for each single case.

Overall, with the cautions we underlined above, our work show that it
is indeed possible to obtain precise ($\langle
\frac{\sigma_z}{1+z} \rangle$ $<$ 0.05) and robust photometric redshifts for more than 90
\% of the galaxies with RIZ $<$ 24.5, in case $u$-band observations are available. In the case of $griz$ ground-based
observations, on the other hand, to get to the required precision we need to eliminate at least $15$ \% of the sample (See Fig. \ref{odds_sel}),
depending on the depth of the survey.

\section*{Acknowledgments}

We acknowledge financial contributions from contracts ASI-INAF
I/023/05/0, ASI-INAF I/088/06/0, ASI I/016/07/0 COFIS, and ASI
EUCLID-DUNE I/064/08/0, ASI-Uni Bologna-Astronomy Dept. Euclid-NIS
I/039/10/0, and PRIN MIUR Dark energy and cosmology with large galaxy surveys. This work was sustained in part by the German Academic Exchange Service (DAAD) under the Vigoni programme. We thank G. Zamorani, H. Hoekstra, F. Abdalla, L. Guzzo, W. Percival and T. Kitching for the useful discussions.

\appendix

\section{Correction to BPZ algorithm}

In Section 3.2 of \citet{bpz}, the author introduces the normalisation parameter $a$ of the template SED as a nuisance parameter in the Bayesian formalism. The probability of observing the colours $C$ given the template $T$ and the redshift $z$ is obtained by integrating the $\chi^2$ likelihood (see Eq . \ref{like}) over $a$, instead of considering only the best-fit value $a_{min}$ as in the simple maximum-likelihood approach. Thus, we have
\begin{equation}
p(C|z,T) \propto \int da \, p(a|m_0) \, p(C|z,T,a),
\end{equation}
where $m_0$ is the magnitude value used as a prior. The prior of $a$ with respect of $m_0$, $p(a|m_0)$, is conservatively assumed flat. Evaluating the integral, he gets
\begin{equation}\label{bpz12}
p(C|z,T) \propto F_{TT}^{-1/2} \, \exp \, \displaystyle \bigg[- \frac{\chi^2(z,T,a_{min})}{2} \bigg],
\end{equation}
where
\begin{equation}\label{bpz13}
F_{TT} =\displaystyle \sum_\alpha \frac{f_{T\alpha}^2 }{\sigma_{f_\alpha}^2}.
\end{equation}
Eq. \ref{bpz12} and \ref{bpz13} mean that the value of $p(C|z,T)$ depends over the ratios of the expected fluxes over the observed errors in the different bands and thus over the input normalisation of the templates (i.d., their value with $a$ = 1). Without any indication on the way one should normalise $f_{T\alpha}$, this introduces an arbitrary factor in the calculation.

To overcome this problem, one can repeat the calculations assuming this time that the template is re-normalised at its best-fit parameter. We define our new templates $f^\prime_{T\alpha} \equiv a_{min} f_T$ and we introduce the nuisance parameter $a^\prime \equiv a/a_{min}$. In this way, we avoid any arbitrariness in the calculation, by defining the templates and the nuisance parameter in a physically motivated way. With the new definitions, Eq. \ref{bpz12} becomes
\begin{equation}\label{pczt}
p(C|z,T) \propto F_{TT}^{\prime{-1/2}} \, \exp \, \displaystyle\bigg[- \frac{\chi^2(z,T,a^\prime_{min} = 1)}{2} \bigg],
\end{equation}
where $F_{TT}^\prime$ is the analogue of $F_{TT}$ with the new template, and the value of $a^\prime$ that
maximizes the likelihood is unity by definition. We can now re-express Eq. \ref{pczt} in terms of the original templates. We have
\begin{equation}\label{ftt}
\sqrt{F_{TT}^\prime} = \bigg(\displaystyle \sum_\alpha \frac{f_{T\alpha}^{\prime2} }{\sigma_{f_\alpha}^2}\bigg)^{1/2} = \bigg(\displaystyle \sum_\alpha \frac{a^2_{min}f_{T\alpha}^2 }{\sigma_{f_\alpha}^2}\bigg)^{1/2}.
\end{equation}
We now make use of Eq. 8 of \citet{bpz}
\begin{equation}\label{bpz8}
a_{min} = \displaystyle \frac{F_{OT}}{F_{TT}},
\end{equation}
where
\begin{equation}
F_{OT} =\displaystyle \sum_\alpha \frac{f_{T\alpha}f_{\alpha} }{\sigma_{f_\alpha}^2}.
\end{equation}
Inserting Eqs. \ref{bpz8} and \ref{ftt} in Eq. \ref{pczt}, and using Eq. \ref{bpz13}, we get
\begin{equation}\label{pczt2}
p(C|z,T) \propto \displaystyle \frac{\sqrt {F_{TT}}}{F_{OT}} \, \exp \, \displaystyle \bigg[- \frac{\chi^2(z,T,a_{min})}{2} \bigg].
\end{equation}
Now, in the term $\sqrt{{F_{TT}}}/{F_{OT}}$ the input normalisation of the original template cancels out, as we desire. Moreover, we can rewrite (see Eq. 7 of \citet{bpz})
\begin{equation}\label{chi2}
\chi^2(z,T,a_{min}) =  \sum_\alpha \frac{f_\alpha^2 }{\sigma_{f_\alpha}^2} - \frac{F_{OT}^2}{F_{TT}}.
\end{equation}
By comparing Eq. \ref{pczt2} and Eq. \ref{chi2} we see that, given the data, the errors and the $\chi^2$ goodness of fit with an unknown template, we have a certain determination of the probability integrated over the normalisation parameter of the template.

In practice, we have modified the BPZ program by inserting Eq. \ref{pczt2} in the place of Eq. \ref{bpz12}. This change eliminated some issues we had expecially with late-type galaxies at high redshift, reducing by more than a factor of two the number of catastrophic failures. We note here that \citet{coe} preferred to eliminate the $F_{TT}^{-1/2}$ term in \ref{bpz12}, reporting numerical instabilities. Actually this would give results very similar to ours, because the ${\sqrt{F_{TT}}}/{F_{OT}}$ term in Eq. \ref{pczt2} is almost irrelevant with comparison to the maximum-likelihood exponential.

\label{lastpage}


\begin{thebibliography}{99}
\bibitem[\protect\citeauthoryear{Abbott et al.}{2005}]{des} Abbott T., Aldering G., Annis J., et al., 2005, arXiv preprint, astro-ph/0510346
\bibitem[\protect\citeauthoryear{Abdalla et al.}{2008}]{abdalla} Abdalla F. B., Amara A., Capak P., Cypriano E. S., Lahav O., Rhodes J., 2008, MNRAS, 387, 969
\bibitem[\protect\citeauthoryear{Abdalla et al.}{2011}]{abdalla2} Abdalla F. B., Banerji M., Lahav O., Rashkov V., 2011, MNRAS, 417, 1891
\bibitem[\protect\citeauthoryear{Abell et al.}{2009}]{lsst} Abell P. A., Allison J., Anderson S. F., et al., 2009, arXiv preprint, astro-ph/0912.0201L
\bibitem[\protect\citeauthoryear{Beckwith et al.}{2006}]{hudf} Beckwith S. V. W., Stiavelli M., Koekemoer A. M., et al., 2006, AJ, 132, 1729
\bibitem[\protect\citeauthoryear{Benitez}{2000}]{bpz} Benitez N., 2000,
ApJ, 536, 571
\bibitem[\protect\citeauthoryear{Bertin \& Arnouts}{1996}]{sextractor} Bertin E., Arnouts S., 1996, A\&AS, 117, 393
\bibitem[\protect\citeauthoryear{Bordoloi et al.}{2010}]{bordoloi}
  Bordoloi R., Lilly S. J., Amara A., 2010, MNRAS, 406, 881
\bibitem[\protect\citeauthoryear{Bolzonella et al.}{2000}]{bolzonella}
  Bolzonella M., Miralles J.-M., Pello R., 2000, A\&A, 363,476
\bibitem[\protect\citeauthoryear{Coe et al.}{2006}]{coe} Coe D., Benitez N., Sanchez S. F., Jee M., Bouwens R., Ford H., 2006, ApJ, 132, 926
\bibitem[\protect\citeauthoryear{Collister \& Lahav}{2004}]{annz}
  Collister A. A., Lahav O., 2004, PASP, 116, 345
\bibitem[\protect\citeauthoryear{Dahlen et al.}{2010}]{dahlen} Dahlen
  T., Mobasher B., Dickinson M., et al., 2010, ApJ, 724, 425
\bibitem[\protect\citeauthoryear{Els et al.}{2009}]{des_see} Els S. G., Schoeck M., Bustos E., et al., 2009, PASP, 121, 922
\bibitem[\protect\citeauthoryear{Erben et al.}{2009}]{erben} Erben T., Hildebrandt H., Lerchster M., 2009, A\&A, 493, 1197
\bibitem[\protect\citeauthoryear{Feldmann et al.}{2006}]{feldmann}
  Feldmann R., Carollo C. M., Porciani C., 2006, MNRAS, 372, 565
\bibitem[\protect\citeauthoryear{Hildebrandt et al.}{2008}]{hilde} Hildebrandt H., Wolf C., Benitez N., 2008, A\&A, 480, 703
\bibitem[\protect\citeauthoryear{Kaiser et al.}{2010}]{panstarrs} Kaiser N. et al., 2002, in Tyson J. A., Wolff S., eds, Proc. SPIE Vol. 4836, Survey and Other Telescope Technologies and Discoveries. SPIE, Bellingham, p. 154
\bibitem[\protect\citeauthoryear{Kitching et al.}{2011}]{great10} Kitching T. et al., 2011, Annals Appl. Stat., 5, 2231
\bibitem[\protect\citeauthoryear{Kuijken et al.}{2008}]{kuijken} Kuijken K., 2008, A\&A, 482, 1053
\bibitem[\protect\citeauthoryear{Laureijs}{2009}]{euclid} Laureijs R., 2009, arXiv preprint, astro-ph/0912.0914
\bibitem[\protect\citeauthoryear{Magnier et al.}{2007}]{pan1} Magnier E., 2007, in Sterken C., ed., ASP Conf. Ser. Vol. 364, The Future of Photometric, Spectrophotometric and Polarimetric Standardization. Astron. Soc. Pac., San Francisco, p. 153
\bibitem[\protect\citeauthoryear{Mei et al.}{2006}]{bpz1} Mei S., Holden B. P., Blakeslee J. P., et al., 2006, ApJ, 644, 759
\bibitem[\protect\citeauthoryear{Melchior et al.}{2007}]{shapelets} Melchior P., Meneghetti M., Bartelmann M., 2007, A\&A, 463, 1215
\bibitem[\protect\citeauthoryear{Menanteau et al.}{2006}]{bpz2} Menanteau F., Ford H. C., Motta V., Benitez N., Martel A. R., Blaxeslee J. P., Infante L., 2006, AJ, 131, 208
\bibitem[\protect\citeauthoryear{Meneghetti et al.}{2008}]{skylens} Meneghetti M., Melchior P., Grazian A., et al., 2008, A\&A, 482, 403
\bibitem[\protect\citeauthoryear{Meneghetti et al.}{2010}]{skylens2} Meneghetti M., Rasia E., Merten J., Bellagamba F., Eltori S., Mazzotta P., Dolag K., Marri S.,  et al., 2010, A\&A, 514, 93
\bibitem[\protect\citeauthoryear{Rafelski et al.}{2009}]{colorpro2} Rafelski M., Wolfe A. M., Cooke J.,  Chen H. W., Armandropf T. E., With G. D., et al., 2009, ApJ, 703, 2033
\bibitem[\protect\citeauthoryear{Rowan-Robinson et al.}{2008}]{swire} Rowan-Robinson M., Babbedge T., Oliver S., et al., 2008, MNRAS, 386, 697
\bibitem[\protect\citeauthoryear{Schoeck et al.}{2009}]{pan_see} Schoeck M., Els S. G., Riddle R., et al., 2009, PASP, 121, 384
\bibitem[\protect\citeauthoryear{Tokovinin \& Travouillon}{2006}]{lsst_see} Tokovinin A., Travouillon T., 2006, MNRAS, 365, 1235
\bibitem[\protect\citeauthoryear{Umetsu et al.}{2010}]{colorpro1}
  Umetsu K., Medezinski E., Broadhurst T., Zitrin A., Okabe N., Hsieh B. C., Molnar S. H., 2010, ApJ, 714, 1470
\bibitem[\protect\citeauthoryear{Way \& Srivastava}{2006}]{way} Way
  M. J., Srivastava A. N., 2006, ApJ, 647, 102
\bibitem[\protect\citeauthoryear{Yee}{1998}]{yee} Yee H. K. C., arXiv preprint, astro-ph/9809347

\end{thebibliography}
\end{document}